\pdfoutput=1

\documentclass[10pt,journal,compsoc]{IEEEtran}
\usepackage[T1]{fontenc}
\usepackage[utf8]{inputenc}
\usepackage{url}
\usepackage{graphicx}
\DeclareGraphicsExtensions{.pdf,.png,.jpg}
\graphicspath{{./img/}}
\usepackage{algpseudocode}
\usepackage{flushend}
\usepackage{amsmath}
\usepackage{float}
\usepackage{amssymb}
\usepackage{glossaries}
\usepackage[super]{nth}
\usepackage{mathptmx}
\usepackage{todonotes}
\newcommand{\coe}{$CO_{2}e$}

\newcommand{\eqtopmargin}{-0.1cm}
\newcommand{\eqbottommargin}{-0.3cm}
\newcommand{\eqtopmarginbig}{-0.4cm}
\newcommand{\eqbottommarginbig}{-0.4cm}
\newcommand{\tabletopmargin}{-0.3cm}
\newcommand{\tablecaptionmargin}{-0.2cm}
\newcommand{\tablebottommargin}{-0.1cm}
\newcommand{\figtopmargin}{-0.2cm}
\newcommand{\figbottommargin}{-0.5cm}
\newcommand{\figcaptionmargin}{-0.7cm}

\newacronym{coe}{\coe}{$CO_{2}$ equivalent}
\newacronym{ewma}{EWMA}{exponentially weighted moving average}
\newacronym{dvfs}{DVFS}{dynamic voltage \& frequency scaling}
\newacronym{vm}{VM}{virtual machine}
\newacronym{era}{ERA}{energy reduction assets}
\newacronym{api}{API}{application programming interface}
\newacronym{os}{OS}{operating system}
\newacronym{rtp}{RTP}{real-time pricing}
\newacronym{qos}{QoS}{quality of service}
\newacronym{sla}{SLA}{service level agreement}
\newacronym{rtep}{RTEP}{real-time electricity pricing}
\newacronym{iaas}{IaaS}{infrastructure as a service}
\newacronym{pm}{PM}{physical machine}
\newacronym{pue}{PUE}{power usage efficiency}
\newacronym{cue}{CUE}{carbon usage effectiveness}
\newacronym{cef}{CEF}{carbon emission factor}
\newacronym{hpc}{HPC}{high-performance computing}
\newacronym{db}{DB}{database}
\newacronym{dc}{DC}{data center}
\newacronym{oltp}{OLTP}{online transaction processing}
\newacronym{mse}{MSE}{mean squared error}
\newacronym{ga}{GA}{genetic algorithm}
\newacronym{arima}{ARIMA}{autoregressive integrated moving average}
\newacronym{ses}{SES}{simple exponential smoothing}
\newacronym{bcf}{BCF}{best cost fit}
\newacronym{bfd}{BFD}{best fit decreasing}
\newacronym{iot}{IoT}{Internet of Things}

\usepackage{tikz}

\usepackage{booktabs}
\usepackage{algpseudocode}
\usepackage{algorithm}
\usepackage{varwidth}
\usepackage{relsize}

\ifCLASSOPTIONcompsoc
  \usepackage[nocompress]{cite}
\else
  \usepackage{cite}
\fi
\ifCLASSINFOpdf
\else
\fi
\begin{document}
%
\title{Pervasive Cloud Controller for \\Geotemporal Inputs}

\author{
	Dražen~Lučanin
	and~Ivona~Brandic%
	\thanks{D. Lučanin and I. Brandic are with the Vienna University of Technology.\hfill\break
    E-mail: \{drazen.lucanin, ivona.brandic\}@tuwien.ac.at}
    \thanks{Manuscript received November 30, 2014.; revised August 3, 2015.}
}

%
%

\markboth{IEEE Transactions on Cloud Computing}%
{Shell \MakeLowercase{\textit{et al.}}: Bare Demo of IEEEtran.cls for Computer Society Journals}
%



\newcommand{\ensavingsnoconsolidation}{$20\%$}
\newcommand{\ensavingsnogeotemp}{$28.6\%$}
\newcommand{\vmnumsimulation}{10k}


\IEEEtitleabstractindextext{%
\begin{abstract}
The rapid cloud computing growth has turned data center energy consumption\
into a global problem.\
At the same time,\ 
modern cloud providers operate multiple geographically-distributed\
data centers.\
Distributed data center infrastructure changes\
the rules of cloud control, as\
energy costs depend on current regional electricity prices and temperatures.\
Furthermore, to account for emerging technologies surrounding\
the cloud ecosystem, 
a maintainable control solution needs to be forward-compatible.\
Existing cloud controllers are focused on \gls{vm} consolidation methods\
suitable only for a single data center\
or consider migration just in case of workload peaks,\
not accounting for all the aspects of geographically distributed data centers.\
In this paper, we propose a pervasive cloud controller\
for dynamic resource reallocation adapting to\
volatile time- and location-dependent factors, \
while considering the QoS impact of too frequent migrations\
and the data quality limits of time series forecasting methods.\
The controller is designed with extensible decision support components.\
We evaluate it in a simulation\
using historical traces of electricity prices and temperatures.\
By optimising for these additional factors,\
we estimate \ensavingsnogeotemp{} energy cost savings\
compared to baseline dynamic \gls{vm} consolidation.\ 
We provide a range of guidelines for cloud providers,\
showing the environment conditions necessary to achieve\
significant cost savings and we validate the controller’s extensibility.\

\end{abstract}

}

\maketitle

\IEEEdisplaynontitleabstractindextext

%
\IEEEpeerreviewmaketitle

\IEEEraisesectionheading{\section{Introduction}\label{sec:introduction}}

%
%
%
%
%
%
%

\IEEEPARstart{T}{o} satisfy growing cloud computing demands,\
data centers are consuming more and more energy,\
accounting for 1.5\%\
of global electricity usage \cite{jonathan_koomey_growth_2011}\
and annual electricity bills of over \$40M\
for large cloud providers \cite{qureshi_cutting_2009}.\
At the same time, a trend of more geographically distributed data centers\
can also be seen, e.g. Google has twelve data centers\
across four continents.\
As new paradigms develop, such as smart buildings\
with integrated data centers \cite{privat_smart_2013},\
computation is shaping as a distributed utility.
Such cloud deployments result in dynamically changing energy cost conditions\ 
and require new approaches to cloud control.

Assorted technological innovations have brought forth\
the optimisation of several independent systems\
that affect cloud operation, creating a heterogeneous\
and dynamically variable environment.\
The technologies of the next-generation electricity grid\
known as the “smart grid”, distributed power generation,\
microgrids and deregulated electricity markets have lead to\
demand response and \gls{rtep} options\
where prices change hourly or even minutely\
\cite{yang_integrating_2013,weron_modeling_2006}.\
Additionally, new solutions for cooling data centers\
(an energy overhead reported to range from 15\% to 45\%\
of a data center's power consumption \cite{barroso_datacenter_2009})\
based on outside air economizer technology result\
in cooling efficiency depending on local weather conditions.\
We call such time- and location-dependent factors geotemporal inputs.\ 
Geotemporal inputs may also include renewable energy availability\
\cite{goiri_parasol_2013},\
peak load electricity pricing \cite{le_reducing_2011}\
or demand response \cite{liu_data_2013,berl_modelling_2013}\
as they constitute time- and location-dependent factors that impact\
the final energy costs as well.\

Furthermore, as IT-based optimisation solutions enter more and more domains,\
we may expect the emergence of new geotemporal inputs in the future.\
Examples include more options for precisely calibrating electricity usage\
and pricing in smart grids, local renewable energy generation,\
further geographical distribution and bringing data centers closer to users\
through smart buildings \cite{privat_smart_2013}\
and all in all more advanced metering infrastructure\
for quantifying cloud service demand and usage through smart cities,\
smart homes, mobile technology or more generally the\
\gls{iot}.\
We denote geotemporal inputs, cloud requirements, regulations and other\
factors that guide the cloud provider's actions\
as decision support components.\ 
We define forward compatiblility as being able to cope with\ 
additional decision support components without drastic changes of\
the core architecture.\
Hence, to account for cloud environment evolution,\
a forward-compatible cloud controller is necessary\
where the decision support components are extensible\
with yet-to-be-realised geotemporal inputs and other factors.

Cloud control approaches can be classified into three levels:\ 
(1) The first level consists only of initial \gls{vm} placement when\
the user requests it.\
For this class of algorithms, existing solutions\
from the field of grid computing \cite{xu_temperature_2013} or\
network request routing \cite{qureshi_cutting_2009} can be applied,\
where geotemporal inputs are used to determine the best placement target.\
Once placed, however, the \gls{vm}, job or network request is never moved.\
(2) The second level is dynamic \gls{vm} consolidation to apply live \gls{vm}\
migrations and optimally reallocate active \gls{vm}s after requests\
to boot new or delete existing \gls{vm}s arrive.\
Existing cloud controllers that apply VM reallocation\
\cite{feller_snooze:_2012,beloglazov_managing_2013,
maurer_enacting_2011} are focused on a model suitable\
for a single data center, where no energy heterogeneity\
inherent to geotemporal inputs is considered.\
(3) The third level, what we call pervasive control,\
is controlling the cloud's resource allocation dynamically to both\
consolidate resources and adapt to\
volatile geotemporal inputs by utilising more cost-efficient data centers\
through long-term planning facilitated by forecasting\
and the asserted data quality.\
The challenges of pervasive control of clouds according to multiple\
decision criteria, including volatile geotemporal inputs\
are that too frequent \gls{vm} migrations to reallocate\
the cloud’s resource consumption cause downtimes \cite{liu_performance_2011}\
which harm the \gls{qos}.\
Forecasting of geotemporal inputs is necessary\
to find the optimal balance between energy cost saving\
and \gls{vm} migration overhead trade-offs.\
With time series forecasting that enables long-term planning,\
the issue of data quality also\
has to be considered to account for the forecasting accuracy\
and reach. Additionally, designing a controller to enable\
turning off or adding new decision support components is necessary\
to integrate the solution into diverse cloud deployments\
and ensure forward compatibility.
To the best of our knowledge, no existing\
cloud control method addresses these challenges.\ 
which is the goal of our work.\

In this paper we propose a novel pervasive cloud controller\
designed for\ 
resource allocation optimisation\
that efficiently utilises cloud infrastructure,\
accounting for geographical data center distribution\
under geotemporal inputs.\
%
%
Our model of a forward-compatible optimisation engine\ 
supports components for\ 
costs based on geotemporal inputs, \gls{vm} migration overheads,\
\gls{qos} requirements and other inputs to be\ 
composed in a unified optimisation problem specification.\
A schedule of \gls{vm} migrations is planned ahead of time\
in a forecast window.\ 
This allows the controller to minimise energy costs by planning\
over a long-term period such as hours or days,\
while retaining the required \gls{qos},\
i.e. not incurring too frequent VM migrations.\
To assess the application of our\ 
pervasive cloud controller\
in diverse cloud deployments,\
we present a number of guidelines showing\ 
how the effectiveness changes under different geotemporal input patterns,\
geographical distributions and forecast data quality.

As a proof of concept evaluation\ 
we present\ 
an implementation of the pervasive cloud controller\
based on a hybrid genetic algorithm\
for optimising the schedule of \gls{vm} migrations.\
A time-series-based schedule representation is developed for integration\
with geotemporal inputs\
to facilitate\
long-term planning.\
A realistic duration-agnostic model,\
with no a priori \gls{vm} lease duration knowledge assumption,\ 
improves the compatibility with real cloud deployments,\
such as Amazon EC2, Google Compute Engine or private\
OpenStack clouds.\ 
Multiple decision support components including energy cost,\
\gls{qos},\ 
migration overhead and capacity constraints\
are combined into an extensible fitness function, matching\
the forward compatibility requirements.\

We evaluate the pervasive cloud controller\ 
in a large-scale simulation consisting of \vmnumsimulation{} \gls{vm}s\ 
using historical electricity price\
and temperature traces\ 
to show the resulting energy cost savings and \gls{qos} impact.\
Based on our simulation results, energy cost savings\
can be increased up to\ 
\ensavingsnogeotemp{} compared to a baseline scheduling algorithm\
\cite{beloglazov_openstack_2014}\
with dynamic \gls{vm} consolidation.\
Furthermore, we expand the evaluation\
to provide guidelines for cloud providers\
in terms of how different geotemporal input value ranges and\
geographical data center distributions\
affect the method's effectiveness.\
We provide a data quality analysis by evaluating\
the controller under different forecasting errors.\
Finally, we validate the architecture’s extensibility\
by performing the simulation with different subsets\
of decision support components.

In the remainder of this paper, Section~\ref{sec:related} examines\
the related work. We explain the research problem intuitively\
on a real example of geotemporal inputs\
and provide a high-level description\
of our pervasive cloud controller in Section~\ref{sec:approach}.\
The formal specification of the plug-and-play decision support components and\
the optimisation problem specification\
is presented in Section~\ref{sec:model}.\
The proof-of-concept implementation of the forward-compatible\
optimisation engine of the pervasive cloud controller we developed\
is explained in Section~\ref{sec:gascheduler} and\
in Section~\ref{sec:evaluation} we describe the evaluation methodology and\
discuss the results.

\begin{figure*}
\vspace{\figtopmargin}
\includegraphics[width=1.0\textwidth]{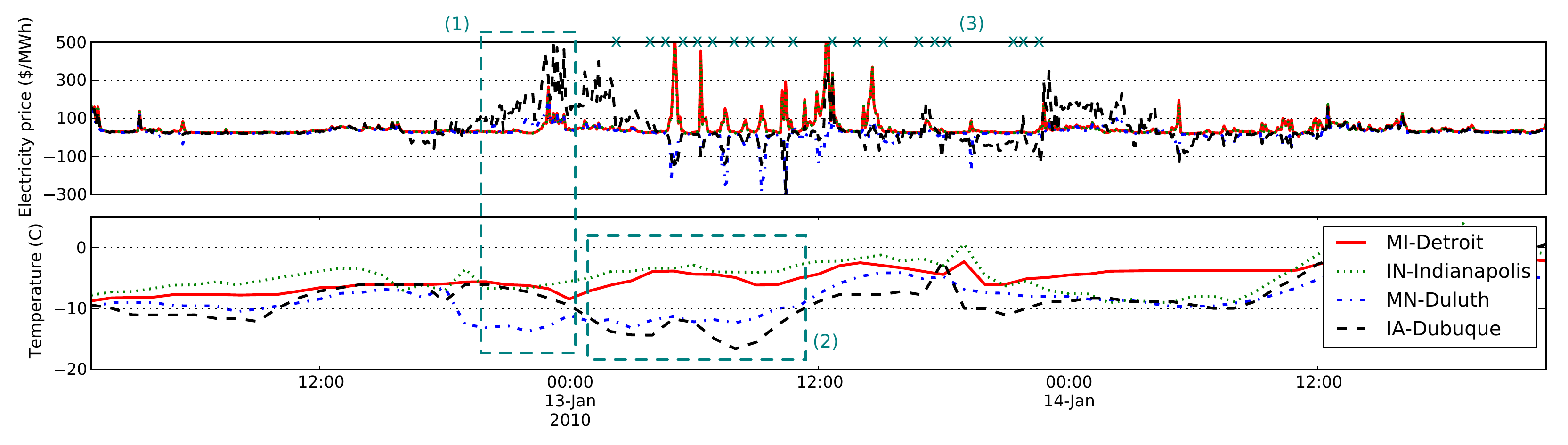}
\vspace{-0.9cm}
\caption{Geotemporal inputs (real-time electricity prices and temperatures)\
at four locations in the USA during a period\
of three days -- temperature values\
within a single day change up to 15 degrees between peaks and lows;\
and even larger relative differences can be observed\
in the volatile electricity prices.\
Electricity price data obtained from \cite{alfeld_toward_2012}\
and temperatures from \cite{_forecast_2015}.}
\label{fig:geotemporal-inputs}
\vspace{\figbottommargin}
\end{figure*}

\section{Related Work}
\label{sec:related}

We structure the related work overview using the already mentioned\
three-level classification of cloud control methods.\
\

Looking at the first level of methods that only perform initial\
placement and consider geotemporal inputs during host selection,\
the approach was pioneered by Quereshi et al. \cite{qureshi_cutting_2009}.\
Their work shows the potential of optimising\
distributed systems (adaptive network routing in content delivery networks)\
for \gls{rtep}, estimating savings up to 40\% of the full electricity cost.\
Similar\ 
routing approaches are explored\
in \cite{rao_minimizing_2010,lin_online_2012,li_towards_2012}\
and considering both electricity prices\
and $CO_2$ emissions in \cite{doyle_stratus:_2013}.\
Initial placement is also researched\
in the context of map-reduce jobs \cite{buchbinder_online_2011}\
and based both on \gls{rtep} and cooling in computational grids\ in \cite{guler_cutting_2013,liu_renewable_2012}.\
A theoretical analysis of placing grid jobs\
with regards to electricity prices, job\
queue lengths and server availability\
is given in \cite{ren_provably-efficient_2012}.
A power-aware job scheduler with no rescheduling\
and assuming a priori job duration knowledge\
is presented in \cite{yang_integrating_2013}.


The second level includes methods targeted\
at modern \gls{iaas} clouds where live \gls{vm} migration is\
used to dynamically reallocate resources and reduce energy consumption.\
These methods, however, value energy the same, no matter\
the time or location, and therefore overlook\
the additional challenges and optimisation potential of geotemporal inputs.\
Feller et al. proposed a distributed scheduling algorithm for dynamic\
\gls{vm} consolidation using live migrations, based on\
hierarchical group management \cite{feller_snooze:_2012}.\
Beloglazov et al. \cite{beloglazov_managing_2013} introduced\
a \gls{vm} consolidation method that minimises the migration frequency\
in an online controller,\
taking future workload predictions into consideration.\
A rule-based \gls{vm} consolidation approach\
is developed in \cite{maurer_enacting_2011}.\
Practical cloud control utilising \gls{vm} migrations\
with a focus on high scalability in production VMware systems\
is researched in \cite{kesavan_practical_2013}.
Consolidation based on RAM and CPU usage is\
researched and evaluated on a real data center\
in \cite{mastroianni_probabilistic_2013}.\

The third level requires pervasive cloud control where \gls{vm}s are\
dynamically migrated to adapt both to user requests and\
changes in geotemporal inputs that enable energy cost savings,\
while considering forecast data quality and \gls{qos} requirements.\
There have been initial advances in this direction.\
Cauwer et al. \cite{cauwer_study_2013} presented a method\
of applying time series forecasting of electricity prices\
to detect how a data center's resource consumption should be controlled in a\
geographically distributed cloud, but for a simplified model with no\
concrete actions that should be applied on \gls{vm}s.\ 
Determining exact per-\gls{vm} actions is a challenging trade-off problem,\
between closely following volatile geotemporal input changes and\
minimising the number of migrations to retain high \gls{qos},\
as we will examine in the following section.\
Abbasi et al. \cite{abbasi_dynamic_2011} started researching migrating\
\gls{vm}s based on \gls{rtep}, but for a limited\
workload distribution scenario where a \gls{pm} hosts only a single \gls{vm}.\
Additionally, temperature-dependent cooling energy overhead and\
forecasting errors are not considered.\
Our work addresses these challenges through a holistic model supporting\
multiple decision support components and\
long-term planning facilitated by forecasting\
and data quality assertion.\

Finally, we look at work related to ours from an algorithmic perspective.\
Approaches for schedule optimisation based on a forecast horizon are\
explored as rolling-horizon real-time task scheduling in\
\cite{zhu_real-time_2014} and a genetic algorithm for this purpose\
was used for multi-airport capacity management\
with receding horizon control in \cite{hu_multiairport_2007}.\
A genetic algorithm for optimising energy consumption in computational grids\
using \gls{dvfs} is presented in \cite{kolodziej_hierarchical_2013}.\
Tabu search, a related meta-heuristic optimisation method,\
is used for static data center location\
and capacity planning\
with a focus on network traffic in \cite{larumbe_tabu_2013}.

\section{Geotemporal Cloud Environments}
\label{sec:approach}

%

We now explain the geotemporal environment surrounding\
geographically-distributed clouds on a real example of electricity\
prices and temperatures. On this use case, we will give an overview\
of the problem on an intuitive level and give a high-level description\
of our pervasive cloud controller, before detailing the\
formal specification of the model in the following section.

\subsection{Problem Overview}

Example geotemporal inputs\
for four US cities\
are shown in Fig.~\ref{fig:geotemporal-inputs}.\
%
%
Rapid changes in geotemporal inputs can occur dynamically.\ 
The peak that can be seen in Dubuque\
on January 12th from 21:00 to 23:00 (1),\
results in five or more times the average prices.\
It can be observed that temperature peaks occur towards the end of the day,\
while lows occur during nights.\
Even though electricity price is more volatile,\
%
partial dependence on previous data points can be seen.\
This means that it is possible to model their behaviour to\
forecast probable future values, and in fact is done\
in practice \cite{weron_modeling_2006,_forecast_2015}.

To explain the potential and challenges of geotemporal inputs\
in the context of cloud computing, let us assume that there are\
two data centers -- one in Detroit and\
another one in Dubuque.\
For the data shown in\
Fig.~\ref{fig:geotemporal-inputs} during the period (1),\
it makes sense to run more \gls{vm}s\
in Detroit when temperatures are the same, because electricity\
is more expensive in Dubuque (constantly over 100 \$/MWh, reaching 500 \$/MWh)\
than in Detroit\
(less than 100 \$/MWh).\
However, when it gets 10 C colder in Dubuque three hours later (2)\
and electricity prices become lower than in Detroit, less\
energy would be consumed on cooling there, resulting in lower energy costs,\
so it is better to migrate\
a number of \gls{vm}s from Detroit to Dubuque\
and shift computational load this way.\

A challenge in adapting cloud control for geotemporal inputs is that\
the cloud provider cannot migrate \gls{vm}s\
between different locations too rapidly,\
as this wastes bandwidth, incurs an energy overhead and impacts \gls{qos}.\
This is underlined even more by the volatile variable behaviour\
observable in electricity prices.\
In Fig.~\ref{fig:geotemporal-inputs},\
we marked by crosses (3)\
all the moments throughout January 13th\
when ratios between electricity prices\
in Detroit and Dubuque change significantly,\
offering an opportunity to save on energy costs by reallocating\
\gls{vm}s using live migrations.\
We can see that 19 migrations would be performed this way.\
If we assume a downtime caused by a live \gls{vm} migration to last for\
one minute, which is possible\
based on the model presented in \cite{liu_performance_2011},\
this would result in a \gls{vm} availability of 98.68\%.\
This availability\
is considerably lower than the 99.95\% availability rates\
advertised by Amazon and Google in their \gls{sla}\ 
and incurs extra data transfer costs.\ 
The challenges arising from this are:\
(1) To profit from geotemporal inputs in cloud computing,\
the trade-offs of the energy savings of geotemporal inputs,\
the migration overheads and impact on \gls{qos}, as well as the data accuracy\
provided by the forecasting methods for future geotemporal inputs all have\
to be considered and reconciled in a long-term plan.\
(2) In reality, the problem has to be\
solved on a much larger scale with more data centers and\
thousands of \gls{vm}s.\
New ways of controlling \gls{vm}s\
across geographically-distributed data centers\
have to be developed to address these challenges.\ 



\subsection{Pervasive Cloud Controller} 

We now present our pervasive cloud controller approach by explaining\
the identified requirements, defining the architecture of the\
solution and giving a high-level overview of its workflow.\ 

\subsubsection{Requirements}
In our approach, we consider\
an \gls{iaas} cloud provider hosted on multiple\
geographically distributed data centers.\
The cloud is assumed to be operating in an environment comprising\
geotemporal inputs such as \gls{rtep} and\
temperature-dependent cooling efficiency\
(other inputs can be added as well).\
The cloud is governed by a controller system\
that manages virtual and physical machines\
in all data centers and can issue actions, such as migrating\
a \gls{vm} from one \gls{pm} to another, suspending or resuming a \gls{pm}.

\vspace{\tabletopmargin}
\begin{table}[H]
\centering
\caption{Example \gls{sla}.}
\vspace{\tablecaptionmargin}
\label{tab:sla} 
\begin{tabular}{lllll}
\toprule
 \#CPUs & RAM & storage & availability & price\\
\midrule
4 & 15 GB & 80 GB & $99.95\%$ & \$0.28/hour\\
\bottomrule
\end{tabular}
\end{table}
\vspace{\tablebottommargin}

The first input are user goals represented by \gls{sla}s, specifying the\
number of requested \gls{vm}s,\
their resource and \gls{qos} requirements.\ 
An example of user requirements for an Amazon \emph{m3.xlarge} \gls{vm}\
specified in an \gls{sla} is shown in Table~\ref{tab:sla}.\
The second input are the geotemporal inputs, providing time series\
metrics describing each of the data center locations, such as\
electricity price and temperature data,\ 
example values of which are shown in Fig.~\ref{fig:geotemporal-inputs}.\
Each geotemporal input is a time series of past and current\
values and, using time series forecasting, it is possible to predict\
future values and the accompanying  data quality\
(reach and the most likely error rate).\
The controller's task is to output a schedule that determines\
where each \gls{vm} is deployed\
and for each \gls{pm} if it is running or suspended at any point in time.\

\subsubsection{Architecture}
\begin{figure}
\centering
\includegraphics[width=1.0\columnwidth]{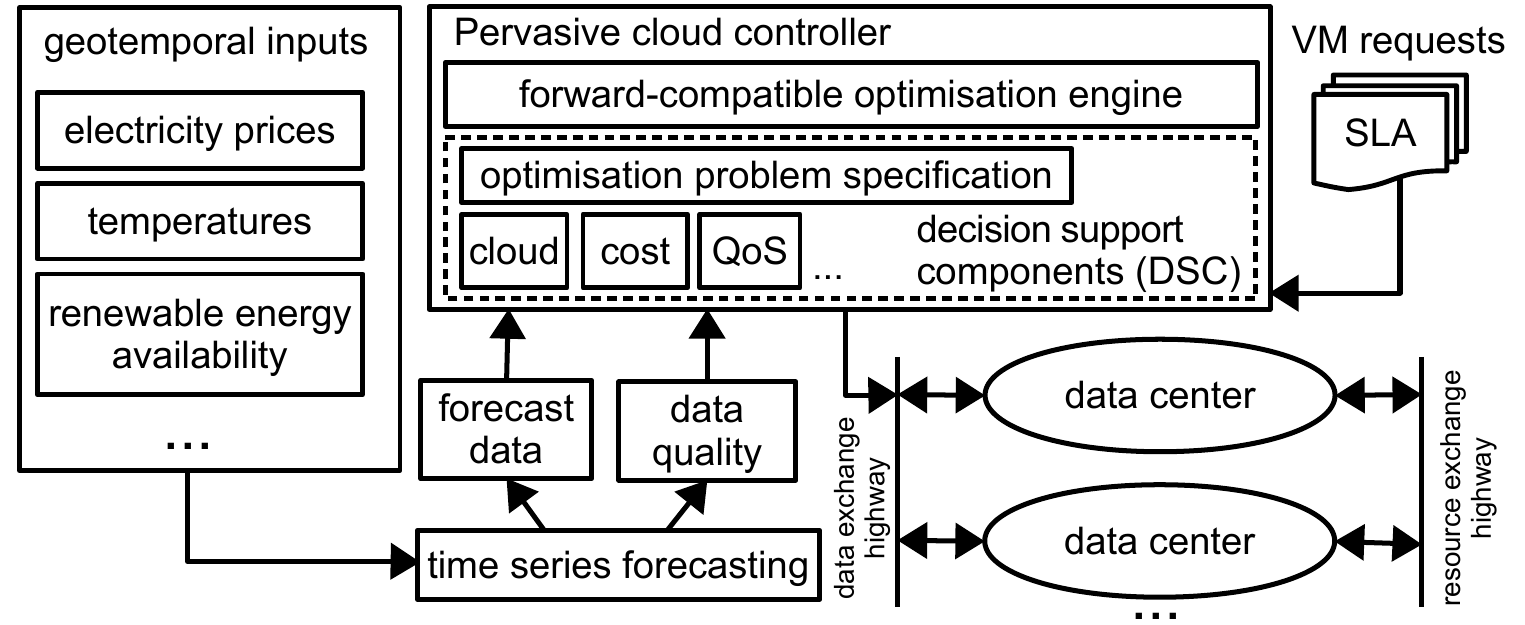}
\vspace{\figcaptionmargin}
\caption{Pervasive cloud controller architecture.}
\label{fig:gascheduler-overview}
\vspace{\figbottommargin}
\end{figure}

Fig.~\ref{fig:gascheduler-overview} shows the architecture of\
our proposed pervasive cloud controller for managing\
a geographically distributed cloud based on geotemporal inputs.\
On a high level, geotemporal inputs are used to obtain forecast\
data and the corresponding quality.\
This input is, together with the \gls{sla}s, provided to the pervasive\
cloud controller, which generates\
a long-term schedule of control actions to apply\
to the geographically distributed data centers.\

Looking at the architecture details,\
geotemporal input forecasts are converted by the controller into\
values meaningful to the cloud provider, e.g. data center energy costs\
that combine \gls{rtep} with the cooling overhead, environmental impact etc.\
These measures along with other internal measures like cloud capacity and\
the \gls{qos} stemming\
from actions planned for \gls{vm}s are all combined\
into an optimisation problem specification as\
decision support components.\
To support new geotemporal inputs, \gls{sla} metrics or cloud regulations,\
it is important for the decision support components to be extensible\
in a plug-and-play manner,\
i.e. without requiring architectural changes.\
We formally present the decision support component model\
in the following section.\
The role of ensuring decision support component extensibility\
lies in a\
forward-compatible optimisation engine.\
It considers the decision support components as criteria\
to plan and optimise\
a schedule of control actions for a future period.\
The challenging part of ensuring forward compatibility\
with new decision support components is that\
there has to be a separation of\
the schedule evaluation logic and the optimisation logic.\
The schedule is evaluated using geotemporal inputs\
and the time-based allocation of \gls{vm}s to \gls{pm}s,\
to estimate the actions' outcome in terms of costs,\
\gls{qos} and any other decision support components.\
This evaluation is then used by the controller in a black-box manner\
to explore the search space of possible actions\
using its custom optimisation logic and the high-level information\
about each schedule returned by the evaluation logic.\
The selected schedule is applied over time\
to physical and virtual machines in the cloud\
(forwarding control actions through a data exchange highway)\
by utilising live \gls{vm} migrations as a resource exchange highway\
to redistribute computational load between the data centers.\



%

\subsubsection{Workflow}
\label{sec:pcc:workflow}

\begin{figure}
\centering
\includegraphics[width=0.7\columnwidth]{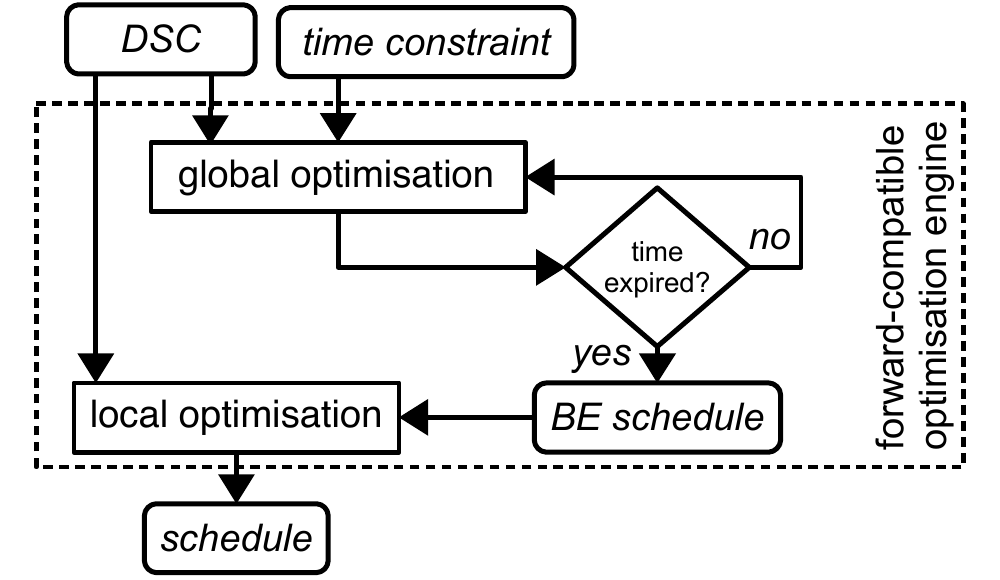}
\vspace{-0.1cm}
\caption{Optimisation engine workflow.}
\label{fig:pcc}
\vspace{\figbottommargin}
\end{figure}

The forward-compatible optimisation engine worfklow\
is illustrated in Fig.~\ref{fig:pcc}.\
It collects the decision support components,\
related together as an optimisation problem specification\
and produces a schedule of\
control actions to apply in the cloud.\
Given that bin packing of allocating \gls{vm}s to \gls{pm}s\
is an NP-hard problem and\
in our case we add to it the dimension of time and time-related\
\gls{qos} requirements (e.g. the \gls{vm} migration frequency),\
an optimal solution can not be found at runtime for an arbitrary\
problem size. To overcome this,\
we propose a two-stage optimisation process.
The first stage is a global optimisation\
method that sweeps the whole search space looking for a global optimum\
within a time constraint (provided as an additional parameter\
that the cloud provider specifies).\
We then take the best-effort schedule this method was able to find\
(BE schedule) and pass it to a second stage local optimisation method that\
continues to improve it with a primary goal\
of satisfying all the hard constraints\
among the decision support components\
that the global optimisation failed to satisfy.\
%
In Section~\ref{sec:gascheduler} we\
present our proof-of-concept implementation\
for the global and local optimisation methods\
based on a hybrid genetic\
algorithm with greedy local constraint satisfaction.\
We later use these methods to evaluate the pervasive cloud controller.\

\section{Decision Support Components}
\label{sec:model}

In this section we formally specify the\ 
cloud, \gls{qos} and cost decision support components\
and show how they are related together\
into an optimisation problem specification\
of optimisation goals and constraints.\
These decision support components are then used by the optimisation\
engine in the schedule generation. 

\subsection{Cloud and \gls{qos} Components}
We consider a single \gls{iaas} cloud provider that is\
represented by $DCs$, a set of $d$ geographically-distributed data centers\
and $PMs$, a set of $p$ physical machines it operates.\
We define each physical machine's location at one of the data centers.

\vspace{\eqtopmarginbig}
\begin{align}
\forall pm \in PMs,\ loc(pm) \in DCs
\end{align}
\vspace{\eqbottommarginbig}

As we are modelling dynamic system behaviour, we define a time period\
in the range from $t_{0}$ to $t_{N}$ (denoted $[t_{0}, t_{N}]$),\
of $N$ discrete (arbitrarily small) periods.
\


User requirements are defined by $vmreqs_t$, a set of virtual machine\
requests at a moment $t$, each of which can either ask for a new \gls{vm}\
to be booted or an existing\
one to be deleted. These events are controlled by end users and we assume\
no prior knowledge of the users' requests\
(as is the case in \gls{iaas} clouds like Amazon EC2).\
Based on the past and current $vmreqs_t$, we define $VMs_t$,\
a set of VMs provided to the users at time $t$.

The cloud provider defines an extensible set of $r$ resource types\ 
that have to be specified through an \gls{sla},\
e.g. number of CPUs and amount of RAM.\
The exact resources for a single \gls{vm} are\
an ordered $r$-tuple of values, defining the \gls{vm}'s $spec$:

\vspace{\eqtopmarginbig}
\begin{align}
spec_{vm} = \left(res_{vm,i},\, \forall i \in \{1, \ldots r\}\right),\ \forall vm \in VMs_t
\end{align}
\vspace{\eqbottommarginbig}

where $res_{vm,i}$ is the $i$-th resource's value.\
For the example shown in Table~\ref{tab:sla}, there are three quantitative\
resource types\
(number of CPUs, amount of RAM and amount of storage)\
that are provided on the infrastructure level, so $r=3$.\
Given the concrete values for an \emph{m3.xlarge} $vm$ instance,\
we have $spec_{vm} = (4, 15, 80)$.\
Similarly, the capacity of a \gls{pm} is defined as an r-tuple\
of the resource amounts it has:

\vspace{\eqtopmarginbig}
\begin{align}
spec_{pm} = \left(cap_{pm,i},\, \forall i \in \{1, \ldots r\}\right),\, \forall pm \in PMs
\end{align}
\vspace{\eqbottommarginbig}

We define the \gls{vm} allocation\ 
at moment $t$ as:

\vspace{\eqtopmarginbig}
\begin{multline}\label{eq:alloc}
\forall pm \in PMs,\ alloc_t(pm) \subseteq VMs_t \\
\text{ s.t. } \forall vm \in alloc_t(pm) \text{ is hosted on $pm$ at moment $t$}
\end{multline}
\vspace{\eqbottommargin}

Effectively, $alloc_t,\ \forall pm \in PMs$\
is the cloud state at moment $t$.\
For any two subsequent moments $t_i$ and\
$t_{i+1},\ t_i \in [t_0, t_{N-1}]$,\
a $vm$ is considered \textit{migrated} if\
$\exists pm_j, pm_k \in PMs$ s.t.\
$vm \in alloc_{t_i}(pm_j)$ and $vm \in alloc_{t_{i+1}}(pm_k)$.\
The number of such migrations for a $vm$ in some relevant period\
specified by the cloud provider\ 
(e.g. an hour) is denoted $R_{mig}(vm)$\ 
and represents the rate of migrations.

\subsection{Cost Components}

Progress has been made in\
modelling various aspects of cloud energy costs and\
we shortly outline the relevant findings\ 
of the existing energy-aware cost model\
using our notation.\
We then proceed with presenting our own\
pervasive cost model\
for expressing energy costs of \gls{iaas} clouds\
based on geotemporal inputs.\

\subsubsection{Energy-Aware Cost Model}

Power consumption $P_t(pm)$ of a $pm \in PMs$\
is modelled in \cite{guler_cutting_2013}\ 
as a function of utilisation $util_t(pm)$ at time $t$,\
with $P_{peak}$ and $P_{idle}$ standing for the\
server's power consumption during peak and idle load, respectively.

\vspace{\eqtopmarginbig}
\begin{align}\label{eq:power_base}
P_t(pm) = pow(util_t, pm) = P_{idle} + util_t(pm) \cdot (P_{peak} - P_{idle})
\end{align}
\vspace{\eqbottommarginbig}

The impact of time series forecasting errors\
is modelled in \cite{cauwer_study_2013},\
where the predicted value $\hat{x_t}$ of a real value $x_t$\
at time $t$ is:\

\vspace{\eqtopmarginbig}
\begin{align}\label{eq:forecasting_error}
\hat{x}_t = \mathcal{N}(x_t, \sigma_{pred}^2),\ \forall t \in [t_{0}, t_{N}]
\end{align}
\vspace{\eqbottommarginbig}

where 
$\mathcal{N}(x_t, \sigma_{pred}^2)$ is
a Gaussian distribution\
with mean $x_t$ and standard deviation $\sigma_{pred}$.\

Temperature-dependent cooling efficiency resulting from\
computer room air conditioning using outside air economizers\
is modelled in \cite{xu_temperature_2013}.\
Cooling efficiency is expressed as partial \gls{pue} $pPUE_{dc,t}$\
at data center $dc$ at time $t$,\
which affects the power overhead based on the following formula:

\vspace{\eqtopmargin}
\begin{align}\label{eq:P_tot}
pPUE_{dc,t} = \frac{P_t(pm) + P_{cool,t}(pm)}{P_t(pm)} = \frac{P_{tot,t}(pm)}{P_t(pm)}
\end{align}
\vspace{\eqbottommargin}

where $P_{cool,t}(pm)$ is the power necessary to cool $pm$, and\
$P_{tot,t}(pm)$ stands for the combined cooling and\
computation power.\
The dynamic value of $pPUE_{dc,t}$ is modelled as a function of\
temperature $T$ to match hardware specifics as:\

\vspace{\eqtopmarginbig}
\begin{align}\label{eq:pPUE}
pPUE_{dc,t} = 7.1705 \cdot 10^{-5}T_{dc,t}^2 + 0.0041T_{dc,t} + 1.0743
\end{align}
\vspace{\eqbottommargin}
%

Based on the migration model developed in \cite{liu_performance_2011},\
the combined energy consumption overhead\
of the source and destination hosts $E_{mig}$\
for a single migration can be calculated as a function of\
the migrated \gls{vm}'s memory $V_{mem}$, data transmission rate $R$,\
memory dirtying rate $D$ and a pre-copying termination threshold $V_{thd}$.

\vspace{\eqtopmarginbig}
\begin{eqnarray}\label{eq:E_mig}
E_{mig} = f(V_{mem}, D, R, V_{thd})
\end{eqnarray}
\vspace{\eqbottommarginbig}

%
%

\subsubsection{Pervasive Cost Model}

Based on our extensible resource types,\
we define a generic model of server utilisation\
$util_t(pm)$ at time $t$ of a $pm \in PMs$\
as a weighted sum\
of the individual resource type utilisations:

\vspace{\eqtopmargin}
\begin{align}
util_t(pm) = \mathlarger{\sum}_{\forall i \in \{1, \ldots r\}} w_i \cdot \frac{\mathlarger{\sum}_{\forall vm \in alloc_t(pm)}res_{vm, i}}{cap_{pm, i}}
\label{eq:util}
\end{align}
\vspace{\eqbottommargin}

where $w_i,\ \forall i \in \{1, \ldots r\}$ is a value in $[0,1]$ describing\
the weight resource type $i$ has on\
the physical machine's power consumption\
(exact amounts depend on hardware specifics; the values we used\
are discussed in Section~\ref{sec:evaluation}).\ 
Variables $cap_{pm, i}$ and $res_{vm, i}$ are the amounts of that resource\
available or requested by the $pm$ or $vm$, respectively.

We model the power consumption $P_t(pm)$ of a $pm \in PMs$\
using the basic approach from Eq.~\ref{eq:power_base}, but\
we extended it to model fast suspension of empty hosts\
(a technology explained in \cite{meisner_powernap:_2009}).\
Also, to model additional load variation,\
we define $P_{peak}$ and $P_{idle}$ as time series of\
a server's power consumption during peak and idle load\
depending on the time $t$, instead of being constant.

\vspace{\eqtopmargin}
\begin{align}\label{eq:power}
P_t(pm) =
\left\{
	\begin{array}{ll}
	    0 & \hspace{-0.5cm} \mbox{if } util_t(pm) = 0\\
		pow(util_t, pm)  & \mbox{otherwise.}
	\end{array}
\right.
\end{align}
\vspace{\eqbottommargin}

%
We use a common time series notation $\{x_t:\ t \in T\}$,\
where $T$ is the index set and $\forall t \in T$, $x_t$\
is the time series value at time stamp $t$.\
For each data center location $dc$ in $DCs$,\
there is a time series of real-time electricity\
prices $\{e_{dc,t}:\ t \in [t_{0}, t_{N}]\}$.\
Similarly, at each location there is a time series of temperature values\
$\{T_{dc,t}:\ t \in [t_{0}, t_{N}]\}$.\
To analyse forecasting errors,\
on both electricity and temperature time series,\
we apply Eq.~\ref{eq:forecasting_error}.\
To explore its impact in the evaluation,\
we vary $\sigma_{pred}$, which determines the accuracy\
of the forecast.\
%
We assumed the temperature-dependent\
cooling efficiency model from Eq.~\ref{eq:P_tot}\ 
to express $P_{tot,t}$
and kept the polynomial model and the fitted\
factors from Eq.~\ref{eq:pPUE}\
where $pPUE$ ranges from 1.02 for -25 C to 1.3 for 35 C.\ 

%

Combining all the equations so far, the cloud's energy cost $C$ can be\
approximated using the rectangle integration method:

\vspace{-0.2cm}
\vspace{\eqtopmargin}
\begin{align}\label{eq:price}
C = \mathlarger{\sum}_{pm \in PMs} \frac{t_N-t_0}{N} \mathlarger{\sum}_{t=t_0}^{t_N-1} P_{tot,t}(pm)e_{loc(pm),t}
\end{align}
\vspace{\eqbottommargin}

Similarly, by omitting the electricity price component $e_{loc(pm),t}$\
we calculate the cloud's energy consumption $E$.\
%
For adding the migration overhead,\
we considered the model\ 
from Eq.~\ref{eq:E_mig}\
and converted it to a cost using a mean electricity price\ 
between the\ 
locations at the time of the migration.\
Bandwidth costs were not considered,\
as the necessary business agreement details are not public\ 
-- e.g. Google leases optical fiber cables,\
instead of paying for traffic.\
The final cost of all the migrations was added to the total energy consumption\
$E$ and total energy cost $C$.

\subsection{Optimisation Problem Specification}

Based on the\ 
decision support components\
we can define the optimisation problem specification as\ 
$\big\{C_i: i \in \{1, \ldots prob\_con\}\big\} \cup \
\big\{G_j: j \in \{1, \ldots prob\_goal\}\big\}$,\
which are the sets of $prob\_con$ constraints and\
$prob\_goal$ optimisation goals composed\
of decision support components.\
This optimisation problem specification can be extended with arbitrary\
requirements.\
We now\ 
state the optimisation problem specification\
with two constraints and two goals\
that we\
use in our evaluation.\

In every moment, every VM has to be allocated to one server\
(belong to its $alloc$ set) that acts as its host.\
This is the \textit{allocation constraint} ($C_1$):

\vspace{\eqtopmargin}
\begin{multline}\label{eq:c1}
\forall t \in [t_0, t_N],\forall vm \in VMs_t,\\
 \exists_{=1} pm \in PMs,\text{ s.t. } vm \in alloc_t(pm)
\end{multline}
\vspace{\eqbottommargin}

The \textit{capacity constraint} ($C_2$) states that at any given time a server cannot host VMs\
that require more resources in sum than it can provide.\

\vspace{\eqtopmargin}
\begin{multline}\label{eq:c2}
\mathlarger{\sum}_{\forall vm \in alloc_t(pm)} res_{vm,i} < cap_{pm,i},\\
\forall pm \in PMs,\ \forall i \in \{1, \ldots r\},\ \forall t \in [t_0, t_N]
\end{multline}
\vspace{\eqbottommargin}

The \textit{cost goal} ($G_1$) is to minimise the cloud's electricity cost $C$\
expressed in Eq.~\ref{eq:price}, stemming from \gls{pm} utilisation,\
cooling efficiency, electricity prices and migration overhead.

The \textit{\gls{qos} goal} ($G_2$) is to minimise the rate of migrations\
$R_{mig}(vm)$ in a designated interval,\
$\forall vm \in VMs_t,\ \forall t \in [t_0, t_N]$.\
In the following section we present an optimisation engine\
for dealing with such a problem specification.\


\section{Forward-Compatible Optimisation Engine}
\label{sec:gascheduler}


In this section we show concrete implementations of the\
optimisation engine workflow from Fig.~\ref{fig:pcc}.\
As already stated in Section~\ref{sec:pcc:workflow},\
to tackle the NP-hard scheduling problem,\
we use a two-stage approach, with best-effort global optimisation\
and a deterministic local optimisation for hard constraint satisfaction.\
For the first stage global optimisation,\
we propose a genetic algorithm \cite{goldberg_genetic_1989} where\
a population of potential solutions\ 
is evolved using\
genetic operators (crossover and mutation).\
For the second stage local optimisation,\ 
we propose a deterministic greedy local search where the best solution\
obtained by the genetic algorithim within the given time limit is\
further improved. The algorithm's main goal is to satisfy the hard\
capacity constraints, in case they were not already satisfied by the\
genetic algorithm, but it also considers the decision support components\
to reduce energy costs based on geotemporal inputs.

\subsection{Algorithm Selection Justification}


The reason the genetic algorithm was chosen for global optimisation\
(in the workflow from Fig.~\ref{fig:pcc}) is that\
using a fitness function for schedule selection\ 
matches the requirement of separated optimisation\
and solution evaluation logic.\
Furthermore, it satisfies\ 
the decision support component extensibility requirement through\
multiple fitness components with associated weights.\ 
There is also a benefit in keeping a population of solutions and not just\
a single best one, as is the case in deterministic optimisation techniques.\
Inputs change over time -- requests to boot new or delete\
old \gls{vm}s arrive,\
temperature or electricity price forecasts change.\
Upon such a change, our genetic algorithm\
propagates\ 
a part of the old\
population to the new environment and there is a higher chance\
that some solutions\ 
will still be fit\
(or\ 
a good evolution basis).\ 

Greedy approaches are often used in deterministic local optimisation,\
e.g. in \cite{beloglazov_energy-aware_2012}.\
For the purpose of improving an existing\
schedule to satisfy primarily the hard constraints,\
without considering the full multi-objective trade-offs,\
it proved as a good addition to the genetic algorithm in our experiments.\

\subsection{Forecast-Based Planning}

It is possible to forecast future values of geotemporal inputs\
to a certain extent \cite{weron_modeling_2006,_forecast_2015}.\
This facilitates planning of more efficient cloud management actions.\
For example, knowing whether a shift in electricity prices between two\
data center locations is the result of a temporary spike or\
a longer trend, enables more cost-efficient scheduling choices.\

Time series forecasting is possible in a\
domain-agnostic manner, by dynamically fitting\
auto-regressive integrated moving average models \cite{melard_automatic_2000}.\
As we are dealing with temperatures and electricity prices, widely used data,\
we assume domain-specific forecast information sources,\
such as the announcement of electricity prices\
(e.g. on day-ahead markets \cite{weron_modeling_2006}\ 
and a weather forecast web service \cite{_forecast_2015}).


\subsection{Cloud Control Schedule}

At the current moment $t_c$, we have information about future values\
for the geotemporal inputs for a period of time\
we call a \textit{forecast window} that ends at $t_f$:\
$fw = [t_c, t_f]$.\
The size of the forecast window is determined by the\
available forecast data and the desired accuracy level.\
Given the current cloud configuration,\ 
we are able to estimate the effects\
of any cloud control actions in terms of the optimisation problem\
inside the forecast window\
by applying the cost model from Eq.~\ref{eq:price}.\
We represent a cloud control schedule as a\
time series $\{action_t:\ t \in fw\}$\
of planned cloud control actions in the forecast window\
.\
In this paper we consider \gls{vm}\
live migration actions \cite{liu_performance_2011}\
and suspension of empty \gls{pm}s that reduces\
idle power consumption \cite{meisner_powernap:_2009}.\
A control action $action_t$\
is described as an ordered pair $(vm,pm)$, specifying\
which $vm$ migrates to which $pm$.\
Migrations determine \gls{vm} allocation over time\
(Eq.~\ref{eq:alloc})\
and implicitly \gls{pm} suspension (Eq.~\ref{eq:power}).

\begin{figure}[H]
\centering
\vspace{\figtopmargin}
\includegraphics[width=1.0\columnwidth]{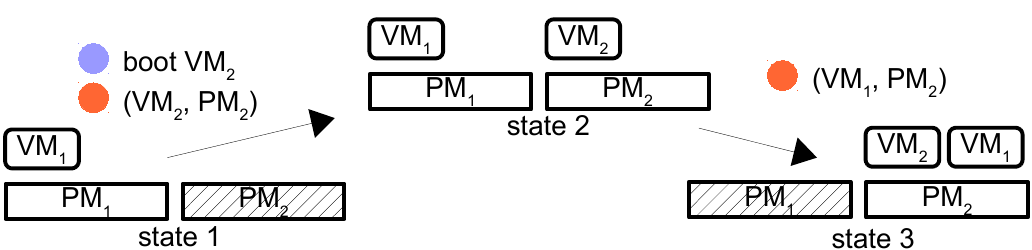}
\vspace{\figcaptionmargin}
\caption{Example of state transitions based on control actions.}
\label{fig:states}
\end{figure}

The representation of the cloud as a sequence of transitions between\
states (as defined through $alloc_t$ in Eq.~\ref{eq:alloc}),\
triggered by migration actions is illustrated in Fig.~\ref{fig:states}.\
Initially, $vm_1$ is hosted on $PM_1$. $PM_2$ is suspended, as it is empty.\
A migration of $VM_2$ to $PM_2$\
transitions the cloud to a new state\
where $PM_2$ is awoken from suspension and hosting $VM_2$.\
An incoming request for \gls{vm} booting\
can be represented as a migration with no source \gls{pm},\
like in the first transition.\
Next, an action migrates $VM_1$ to $PM_2$, after which $PM_2$ is hosting\
both \gls{vm}s and $PM_1$ is suspended.

\begin{figure}[H]
\centering
\vspace{\figtopmargin}
\includegraphics[width=1.0\columnwidth]{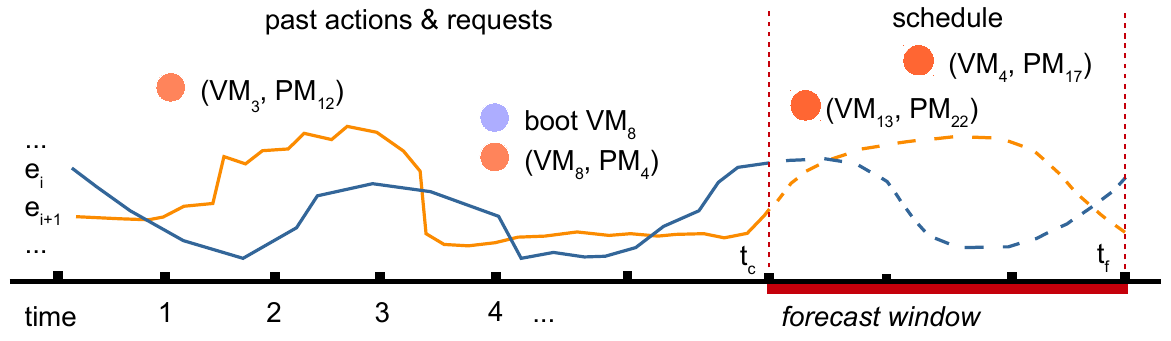}
\vspace{\figcaptionmargin}
\caption{Schedule optimisation inside a forecast window.}
\label{fig:forecasting}
\end{figure}

The time-based aspect of a schedule\
is illustrated in Fig.~\ref{fig:forecasting}.\
Past events that occurred before the current moment $t_c$,\
such as a \gls{vm} migration at $time=1$ or a new user request\
to boot a \gls{vm} at $time=4$, determine the current cloud state.\
Based on the past values of different geotemporal variables, such\
as electricity prices $e_i$ and $e_{i+1}$ at different data center\
locations, we are able to get their value forecasts in the\
forecast window.\
Different control actions of a schedule inside the forecast window\
can then be tried out\
at any moment between $t_c$ and $t_f$.\
The control actions can be evaluated\
to determine the resulting \gls{vm} locations and estimate costs with\
regard to the different geotemporal input forecasts and other optimisation\
problem aspects, such as constraints or \gls{sla} violations.\
When a schedule has been selected for execution, any immediate actions\
are applied and the forecast window moves as time passes,\
new requests arrive and geotemporal inputs change.\

\subsection{Hybrid Genetic Algorithm Implementation}



We now present\ 
the hybrid genetic algorithm.\ 
The most challenging parts in its design were\ 
the genetic operators that had to semantically match\
our optimisation problem domain,\
the fitness function that combines the decision support components\
and the hybrid part of the algorithm, i.e. the greedy\
local improvement.\ 

\subsubsection{Schedule Fitness}

In a genetic algorithm, we keep not only one, but a population of multiple problem\
solutions -- multiple cloud control schedules, in our case.\
An essential part of the algorithm is to evaluate the fitness\ 
of each of these schedules.\
The fitness function we derived for this purpose is adapted from the\
decision support components, but constrained only to the forecast window\
(as we cannot affect actions that were already executed or plan further\
ahead than the available forecasts).\ 
Additional decision support components can be treated in a similar manner.\

For a schedule $s$, we calculate the capacity and\
allocation constraint fitness, as a measure of how well\
the \gls{vm}s are allocated and the \gls{pm}s are within their capacity,\
using the time-weighted function:

\vspace{\eqtopmargin}
\begin{multline}
constraint(s) = \mathlarger{\sum}_{\forall t \in fw}\frac{w_{alloc} \cdot R_{unalloc, t} + w_{cap} \cdot R_{overcap, t}}{|fw|}
\end{multline}
\vspace{\eqbottommargin}

where $R_{unalloc, t}$ and $R_{overcap, t}$ are ratios of\
$vm, \forall vm \in VMs_t$ for which $C_1$ (Eq.~\ref{eq:c1}) does not hold and\
$pm, \forall pm \in PMs$ for which $C_2$ (Eq.~\ref{eq:c2})\
does not hold at moment $t$, respectively.\
$|fw|$ is the cardinality of $fw$\
(the number of time periods in the interval).\
Constants $w_{alloc}$ and $w_{cap}$ in $[0, 1]$ define the components' weights.

The acceptable migration rate $R_{mig\_min}$ and the maximum allowed\
migration rate $R_{mig\_max}$ can be specified in the \gls{sla},\
or determined by the cloud provider's bandwidth expenses.\
Their values are used to calculate the \gls{qos} component\
from the actual migration rate $R_{mig}$ over the period of $fw$.

\vspace{\eqtopmargin}
\begin{align}
qos\_pen(vm) =
\left\{
	\begin{array}{ll}
	    0 & \mbox{if } R_{mig} < R_{mig\_min}\\
   	    1 & \mbox{if } R_{mig} > R_{mig\_max}\\
	    \frac{R_{mig} - R_{mig\_min}}{R_{mig\_max} - R_{mig\_min}} & \mbox{otherwise.}
	\end{array}
\right.
\end{align}
\vspace{\eqbottommargin}

The $qos\_pen$ expression gives a penalty for\
too frequent migrations per \gls{vm}s\
that grows linearly from 0 to 1 for the migration rate $R_{mig}$\
from $R_{mig\_min}$ to $R_{mig\_max}$. We then calculate the average\
\gls{qos} penalty over all the \gls{vm}s.

\vspace{\eqtopmargin}
\begin{align}
qos(s) = \frac{\mathlarger{\sum}_{\forall vm \in VMs_{t_c}}qos\_pen(vm)}{|VMs_{t_c}|}
\end{align}
\vspace{\eqbottommargin}

As a cost estimation, we use a simplified expression -- $utilprice$.\
First, we calculate the average value of utilisation\
multiplied by $pPUE$ and $e$\
over all \gls{pm}s in the forecast window.\
This is a heuristic of the total energy costs from Eq.~\ref{eq:price}.

\vspace{\eqtopmargin}
\begin{multline}
up\_avg(s) = \frac{\mathlarger{\sum}_{\forall pm \in PMs}\mathlarger{\sum}_{\forall t \in fw}util_t(pm) \cdot pPUE_{loc(pm),t} \cdot e_{loc(pm),t}}{|PMs|\cdot|fw|}
\end{multline}
\vspace{\eqbottommargin}

Then we normalise it to a $[0,1]$ interval\
by dividing it with $up\_worst$,\
which is the same as $up\_avg$, but calculated\
for a constant maximum utilisation $util_t(pm)=1$.

\vspace{\eqtopmargin}
\begin{align}
utilprice(s) = \frac{up\_avg(s)}{up\_worst}
\end{align}
\vspace{\eqbottommargin}

To measure how tightly \gls{vm}s are packed among the available \gls{pm}s,\
we estimate the consolidation quality $consolid$\ 
by considering only positive utilisation time series elements denoted as\
$\{ pos\_util_t(pm): t \in fw,\
\text{ s.t. } util_t(pm)>0 \}$.

\begin{align}
consolid(s) = 1 - \frac{1}{|PMs|} \mathlarger{\sum}_{\forall pm \in PMs}\frac{\mathlarger{\sum}_{\forall t \in fw}pos\_util_t(pm)}{|pos\_util(pm)|}
\end{align}
\vspace{\eqbottommargin}


Finally, we can calculate the fitness as:

\vspace{\eqtopmarginbig}
\begin{align}
\label{eq:fitness}
\begin{split}
fitness(s) = w_{ct} \cdot constraint(s) + w_q \cdot qos(s) +\\
+ w_{up} \cdot utilprice(s) + w_{cd} \cdot consolid(s)
\end{split}
\end{align}

with $w_{ct}$, $w_q$, $w_{up}$ and $w_{cd}$ in $[0, 1]$ determining\
each component's impact.\ 
The optimal schedule converges towards a fitness of 0 and the\
worst schedule towards 1.\
This solution evaluation form suits the forward compatibility requirement,\
as it can easily be extended with new decision support components by\
weighing them into the total fitness summation.\

\subsubsection{Genetic Operators}

The \textit{creation} procedure creates a random schedule as\
a time series of actions $(vm,pm),\ vm \in VMs_{t_c}, pm \in PMs$\
at random times $t_r\ \in fw$. The number of migrations is uniformly\
distributed between the parameters $min\_migrations$ and $max\_migrations$.\

Given schedules $s_1$ and $s_2$, the \textit{crossover} operator creates\
a child schedule $s_3$ by choosing a random moment\
$t_r\ \in [t_c, t_f]$:

\vspace{\eqtopmargin}
\begin{align}\label{eq:crossover}
s_3 =
\left\{
	\begin{array}{ll}
	    s_{1,t}: & t \in [t_c, t_r],\\
   	    s_{2,t}: & t \in [t_r, t_f]
	\end{array}
\right\}
\end{align}
\vspace{\eqbottommargin}


The \textit{mutation} operator applied to a schedule $s$ removes a random\
action $a$\ 
and inserts\ 
one at a random moment $t_r\ \in fw$:

\vspace{\eqtopmarginbig}
\begin{multline}\label{eq:mutation}
s = s \setminus a \cup \{t_r : (vm,pm)\},\\
a \in s, vm \in VMs_{t_c}, pm \in PMs
\end{multline}
\vspace{\eqbottommarginbig}

\begin{table*}[ht]
\vspace{\tabletopmargin}
\centering
\caption{Simulation parameters}
\vspace{\tablecaptionmargin}
\label{tab:parameters-simulation} 
\begin{tabular}{ccccccccccccc}
\toprule
 period & duration & d & p (CPU; RAM)    & v (CPU; RAM) & $P_{peak}$ & $P_{peak}$ & $(\mu_P, \sigma_P^2)$ & $R_{mig\_min}$ &   $R_{mig\_max}$ & $R$ & $D$ & $V_{thd}$ \\
\midrule
  1 h   & 2 weeks  & 6 & 2,000 (8-16; 16-32) & 10,000 (1-2; 2-4) & 200 & 100 & (0, 25) &   1/4 &
       1 & 1 Gb/s & 0.3 Gb/s & 0.1 Gb/s \\
\bottomrule
\end{tabular}
\vspace{\tablebottommargin}
\end{table*}

\begin{table*}[ht]
\vspace{\tabletopmargin}
\centering
\caption{Optimisation engine parameters}
\vspace{\tablecaptionmargin}
\label{tab:parameters-scheduler} 
\begin{tabular}{cccccccccccccccc}
\toprule
 fw &   pop &   gen &   cross &   mut &   rand   & $min\_migr$ & $max\_migr$ & $w_{up}$ & $w_{ct}$ & $w_{alloc}$ & $w_{cap}$ & $w_q$ & $w_{cd}$ \\
\midrule
 12 h &   100 &   100 &    0.15 &  0.05 &    0.3  & 0 & $fw |VMs_{t_c}|/3$ & 
 0.4 & 0.1 & 0.4 & 0.6 & 0.4 & 0.1 \\
\bottomrule
\end{tabular}
\vspace{\tablebottommargin}
\end{table*}

\subsubsection{Algorithm}

The core \gls{ga} is listed in Alg.~\ref{alg:ga}.\
Among other parameters not introduced so far, it also receives\
the population size $pop$, crossover, mutation and random creation rates\
$cross$, $mut$ and $rand$ in $[0,1]$ and the\
maximum number of generations $gen$.

\begin{algorithm}
\caption{Core genetic algorithm.}
\label{alg:ga}
\begin{algorithmic}[1]
\State $schedules = create\_population(pop, rand, [existing])$ \label{alg:ga:creation}
\State $i = 0$
\While{$True$}
	\State $calculate\_fitness(schedules)$ \label{alg:ga:fitness} \Comment{Eq.~\ref{eq:fitness}}
	\State sort $schedules$ by fitness, best first
	\If{$i = gen$} \label{alg:ga:termination}
		\State break loop
	\EndIf
	\State $parents = roulette\_wheel(schedules, cross)$
	\State $children = crossover(parents)$  \Comment{Eq.~\ref{eq:crossover}}
	\State $schedules = schedules[0:pop\cdot(1-cross)] + children$\label{alg:ga:new}
	\State $mutation(schedules, mut)$\label{alg:ga:mutation}  \Comment{Eq.~\ref{eq:mutation}}
	\State $i = i + 1$
\EndWhile
\State $existing = schedules$
\State \Return $schedules[0]$
\end{algorithmic}
\end{algorithm}
%


After creating the population, every schedule is evaluated using\
the fitness function (Eq.~\ref{eq:fitness}) in line~\ref{alg:ga:fitness}.\
Schedules are selected for crossover using the \textit{roulette wheel}
selection method with chances for selection\
proportional to a schedule's fitness.\
A new generation is created by replacing the worst schedules with\
the newly created children (line~\ref{alg:ga:new}).\
A random segment of the population proportional to $mut$ is\
changed by applying the mutation operator (line~\ref{alg:ga:mutation}).\
The termination condition is fulfilled after $gen$ generations\
have passed (line~\ref{alg:ga:termination}) and the best schedule\
is returned.\
To cope with forecast-based planning,\
an existing population is partially propagated to the\
next schedule reevaluation, after the forecast window moves (line~\ref{alg:ga:creation}).\


The \gls{ga} is not guaranteed to satisfy all the constraints\
within a limited time period. To remedy this, we expand the optimisation with\
a greedy constraint satisfaction algorithm that is applied to the best\
schedule selected by the genetic algorithm to reallocate any offending\
\gls{vm}s using a \gls{bcf} heuristic we developed\
that also considers geotemporal inputs.\

The greedy constraint satisfaction pseudo-code is listed in Alg.~\ref{alg:bcf}.\
The algorithm receives as input $schedule$, the output of the \gls{ga}.\
We begin by marking for reallocation all \gls{vm}s which cause\
any hard constraint violations ($C_1$ or $C_2$) in line~\ref{alg:bcf:nonalloc}\
and additionally all \gls{vm}s from underutilised hosts\
in line~\ref{alg:bcf:underutil}.\ 
The \gls{vm}s will then be reallocated\
in the outermost loop (line~\ref{alg:bcf:vmloop})\
starting with larger \gls{vm}s first\
whose placement is more constrained (line~\ref{alg:bcf:vmsort}).\
Available \gls{pm}s are split into $active$ and $nonactive$ lists,\
depending on whether they are suspended or not.\
We sort $inactive$ in line~\ref{alg:bcf:sortinactive}\
such that larger \gls{pm}s come first for activation\
(preferable to more smaller machines, because of the idle power overhead)\
and data centers with lower combined electricity price\
and cooling overhead cost are preferred.\
The target \gls{pm} to host $vm$ is selected in the inner loop\
in line~\ref{alg:bcf:mappedloop} by first sorting $active$\
to try and fill out almost full \gls{pm}s first and prefer\
lower-cost location in case of ties (line~\ref{alg:bcf:sortactive})\
and activating the next \gls{pm} from $inactive$ if $vm$ does not\
fit any of the $active$ \gls{pm}s (line~\ref{alg:bcf:popinactive}).\
When a fitting \gls{pm} is found, the action is added\
to $schedule$ and the algorithm continues with the next \gls{vm}.

\begin{algorithm}
\caption{Greedy constraint satisfaction -- \gls{bcf} heuristic.}
\label{alg:bcf}
\begin{algorithmic}[1]
\State $to\_alloc =$ empty list
\State append all constraint-violating \gls{vm}s to $to\_alloc$ \label{alg:bcf:nonalloc}
\State append \gls{vm}s from all underutilised \gls{pm}s to $to\_alloc$ \label{alg:bcf:underutil}
\State sort $to\_alloc$ by resource requirements decreasing \label{alg:bcf:vmsort}
\For{$vm \in to\_alloc$} \label{alg:bcf:vmloop}
    \State $active = $ all PMs where at least one VM is allocated 
    \State $inactive = $ all PMs where no VMs are allocated
    \State sort $inactive$ by capacity decreasing, cost increasing\
    \label{alg:bcf:sortinactive}
    \State $mapped = False$
    \While{not $mapped$} \label{alg:bcf:mappedloop}
        \State sort $active$ by free capacity, cost increasing \label{alg:bcf:sortactive}
        \For {$pm \in active$}\label{alg:bcf:pmloop}
            \If {$vm$ fits $pm$}\label{alg:bcf:pmfits}
                \State $mapped = True$
                \State break loop
            \EndIf
        \EndFor
        \If {not $mapped$}
            \State pop $inactive[0]$ and append it to $active$ \label{alg:bcf:popinactive}
        \EndIf
    \EndWhile
    \State modify $schedule$ by adding a migration $(vm, pm)$
\EndFor
\State \Return $schedule$
\end{algorithmic}
\end{algorithm}

\section{Evaluation}
\label{sec:evaluation}


We evaluated the proposed progressive cloud controller in a large-scale simulation of \vmnumsimulation{} \gls{vm}s based\
on real traces of geotemporal inputs.\
To be able to simulate cloud behaviour under geotemporal inputs,\
we developed our own open source simulator\
Philharmonic.\ 
%
The goals of the evaluation are:\
(1) Estimate energy and cost savings, as well as the \gls{qos}\
attainable using the pervasive cloud controller;\
(2) Analyse the impact of various inputs, such as data center geography,\
different geotemporal inputs or controller parameters;\
(3) Validate the pervasive cloud controller extensibility by\
running the simulator with different decision support component subsets;\
(4) Define cloud provider guidelines, such as how temperature variation\
or forecast data quality affect the energy savings and illustrate their\
usage in a case study.

\subsection{Evaluation Methodology}
\label{sec:evaluation_methodology}

In this part we give an overview of the simulator's implementation\
and proceed with explaining all the simulation details,\
such as the datasets, parameters and the baseline controller.

\subsubsection{Philharmonic Simulator}

\begin{figure}
\vspace{\figtopmargin}
\centering
\includegraphics[width=0.7\columnwidth]{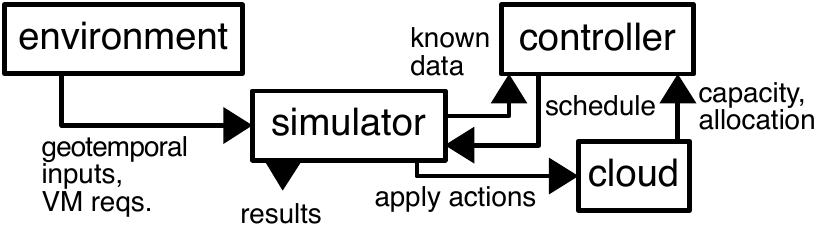}
\vspace{-0.2cm}
\caption{Philharmonic simulator overview.}
\label{fig:philharmonic}
\end{figure}

A high-level overview of\ 
the Philharmonic\
simulator \cite{drazen_lucanin_philharmonic_2014} that we developed\
is shown in Fig.~\ref{fig:philharmonic}.\
A simulation in Philharmonic\ 
consists of iterating through\
a series of equally-spaced time periods,\
collecting the currently available electricity price\
and temperature forecasts, as well as the incoming \gls{vm} requests\
from the environment component.\
The controller is called with the data known at that moment\
about the environment and the cloud\
to reevaluate the schedule for the forecast window and potentially\
schedule new or different actions.\
The simulator applies any actions scheduled for\
the current moment on the cloud model\
and continues with the next time step, repeating\
the procedure.\
The applied actions are used to\
calculate the resulting energy consumption and electricity\
costs, using the model from Section~\ref{sec:model}.

\subsubsection{Simulation Details}

\begin{figure}
\vspace{\figtopmargin}
\centering
\includegraphics[width=1.0\columnwidth]{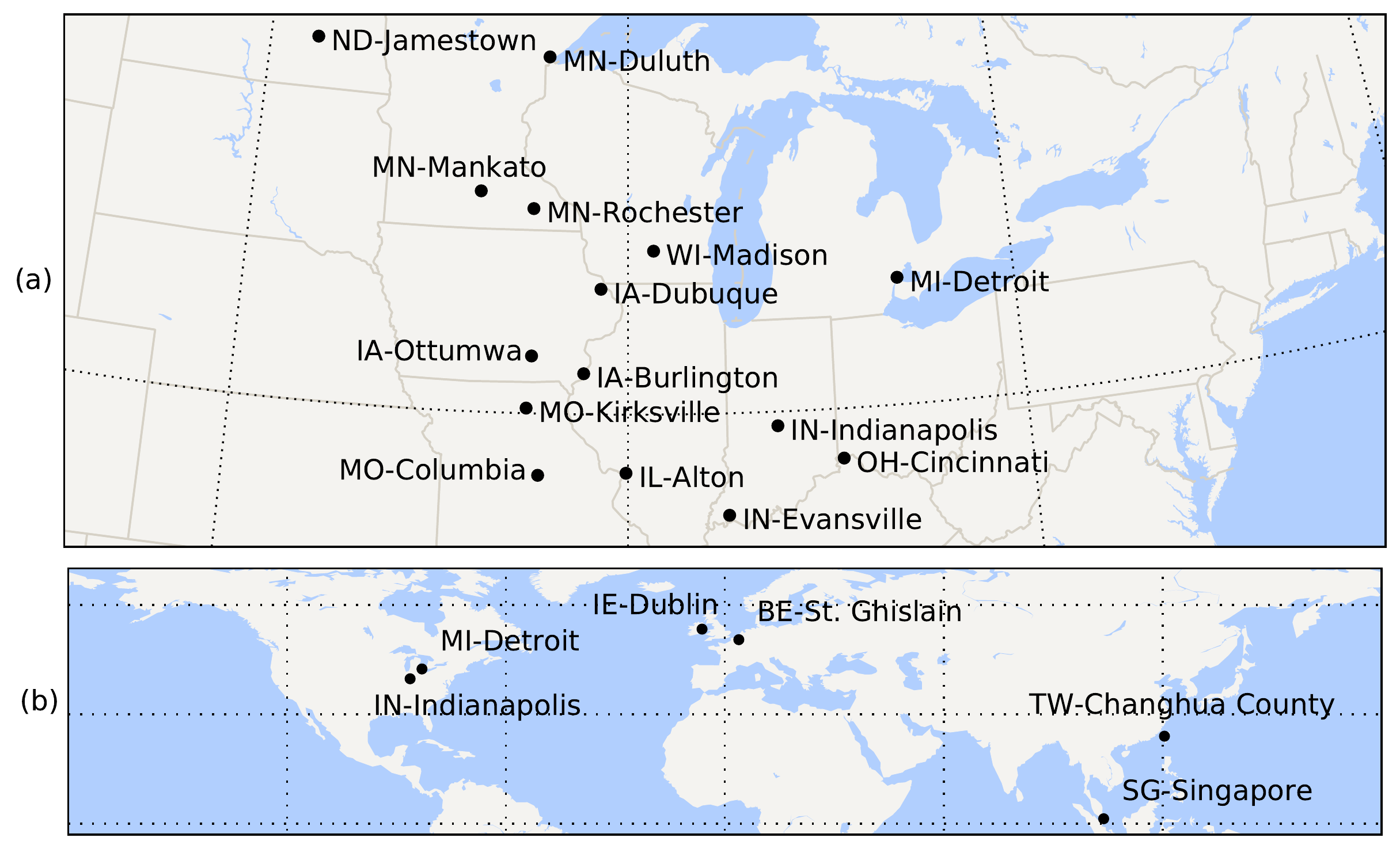}
\vspace{\figcaptionmargin}
\caption{Cities used as data center locations in the simulation based on a:\
(a) US dataset
(b) world-wide dataset.}
\label{fig:cities}
\vspace{\figbottommargin}
\end{figure}


Real historical traces for electricity prices and temperatures\
were used in the simulation for 15 cities in the USA\
shown in Fig.~\ref{fig:cities} (a).\
The electricity price dataset described in \cite{alfeld_toward_2012} was used.\
The temperatures were obtained from\
the Forecast web service \cite{_forecast_2015}.\
To evaluate less correlated geotemporal inputs,\
we used a world-wide dataset for six cities accross three continents\
shown in Fig.~\ref{fig:cities} (b),\
choosing non-US cities to match\
the locations of Google's data centers.\ 
Temperatures were again obtained\
from the Forecast API \cite{_forecast_2015}.\
Due to lack of \gls{rtep} data for the four non-US cities,\
we artificially generated these traces from the data\
known for other US cities.\
We shifted the time series based on the time zone offsets\
and added a difference in annual mean values to resemble\
local electricity prices.

\begin{figure}
\vspace{\figtopmargin}
\includegraphics[width=1.0\columnwidth]{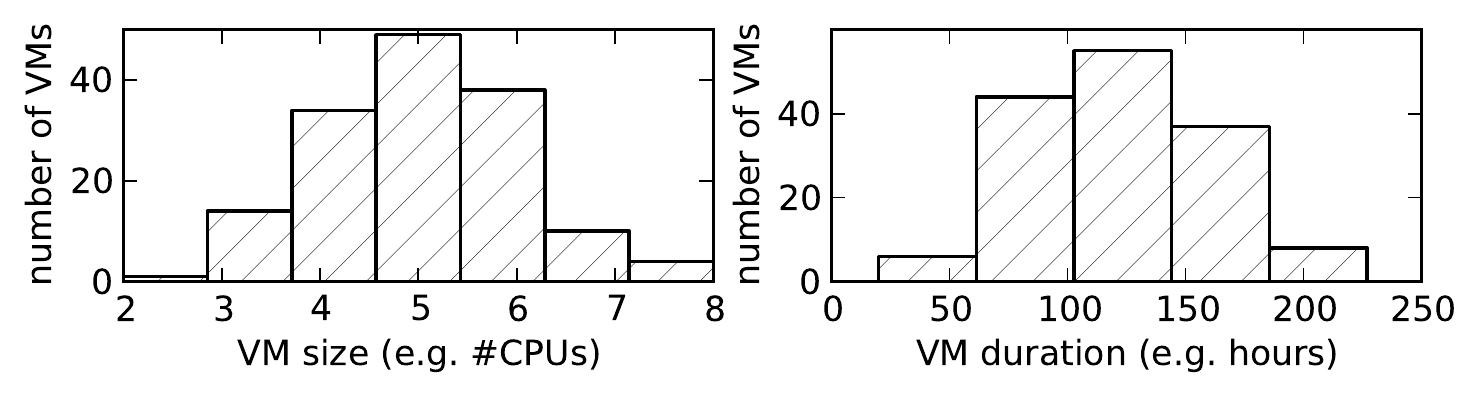}
\vspace{-0.85cm}
\caption{VM request resource and duration histogram.}
\label{fig:vmreqs}
\vspace{\figbottommargin}
\end{figure}

User requests for \gls{vm}s were generated randomly by\
uniformly distributing the creation time and duration.\
The specifications of the
requested \gls{vm}s were modelled\
by normally distributing each resource type.\
An example \gls{vm} request distribution is illustrated\
in Fig.~\ref{fig:vmreqs}.\
%
%
Available cloud infrastructure was generated by uniformly\
distributing physical machines among different data centers.\
Capacities for each each machine's resources were generated\ 
in the same manner as the \gls{vm} requests.

\begin{figure*}
\includegraphics[width=1.0\textwidth]{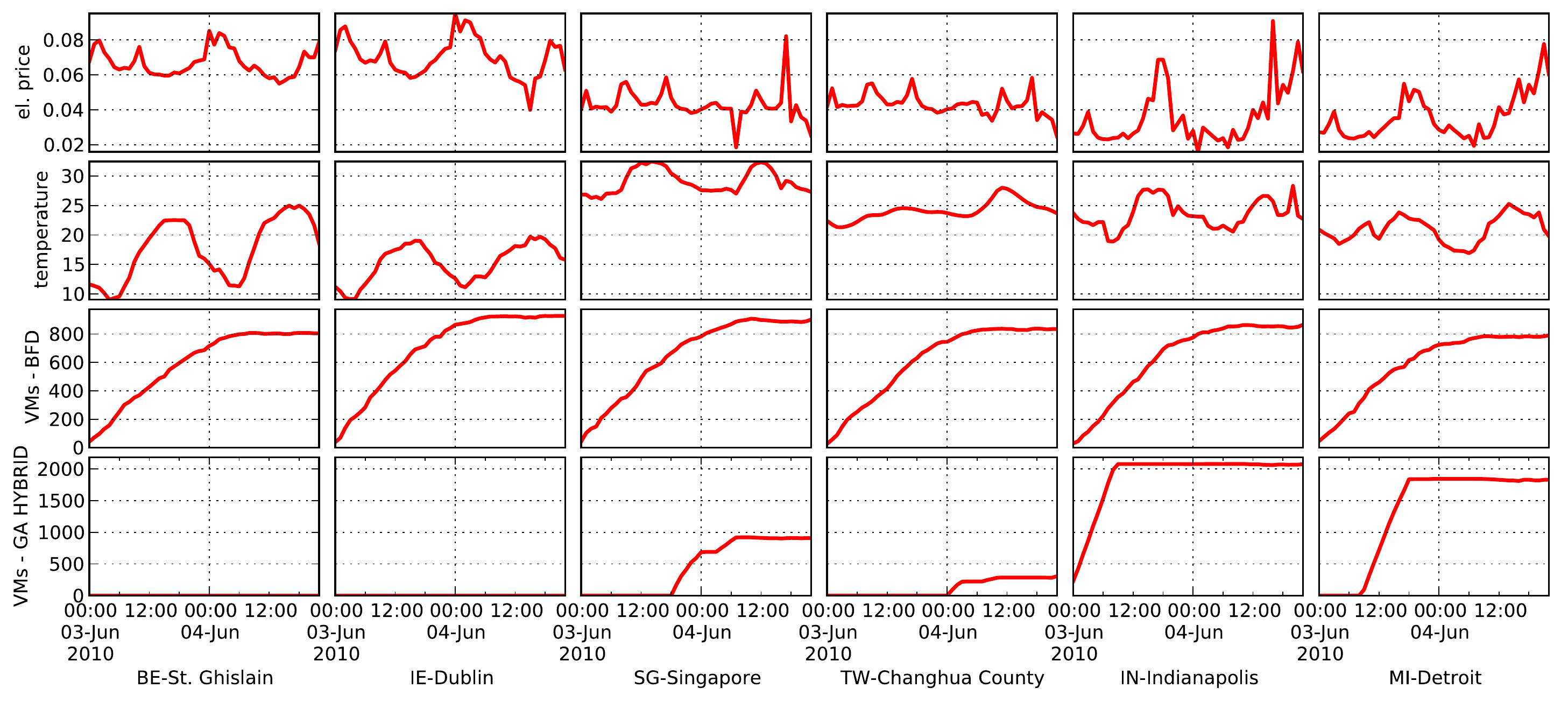}
\vspace{\figcaptionmargin}
\caption{Dynamic \gls{vm} management comparison of the controllers.}
\label{fig:results-dynamic}
\vspace{\figbottommargin}
\end{figure*}

The exact simulation parameters used in the evaluation\
are listed in Table~\ref{tab:parameters-simulation}.\
The time is defined by its period (simulation step size) and\
the total duration that determines the number of steps.\
To define the cloud, the number of data centers is given as $d$\
(for the world-wide scenario, for the US scenario we consider $d=15$),\
and for $PMs$ and $VMs$ their number $p$ and $v$\
(of boot requests in case of \gls{vm}s --\
there were about 50\% as much delete events as well)\
with minimum and maximum resource\
values in parentheses for the resources\
we assumed in this simulation -- number of CPUs\
and the amount of RAM in GB.\
Besides running the simulation for the large-scale scenario with\
\vmnumsimulation{} \gls{vm}s, we also simulated\
100 and 1k \gls{vm}s, but as the difference in normalised savings\
was only marginal, we only include the large-scale results.\
Uniform resource weights were used in the $util$ function (Eq.~\ref{eq:util}).\
We used $P_{peak}$ with $P_{idle}$ values $50\%$ of the peak,\
as reported in \cite{fan_power_2007}, and added normal random noise\
of the form $\mathcal{N}\ (\mu_P, \sigma_P^2)$ to account for load variation.\
$R_{mig}$ was calculated hourly\
and constant migration model parameters ($R$, $D$, $V_{thd}$) were used.\
We used these settings in all the simulations, unless otherwise specified.\



As a baseline controller for results comparison, we implemented a method\
for \gls{vm} consolidation dynamically adapting to user requests using a\
\gls{bfd} placement heuristic developed\
by Beloglazov et al. in \cite{beloglazov_energy-aware_2012}.\
We implemented the updated version of the controller that is currently\
developed for inclusion in\
the OpenStack open source cloud manager\ 
as project Neat \cite{beloglazov_openstack_2014}.\
Based on our classification of related work, this is\
a level two cloud management\
method that dynamically reallocates \gls{vm}s,\
treating all energy uniformly.\


The optimisation engine's algorithm parameters\
are listed in Table~\ref{tab:parameters-scheduler}.\
All the weights used in the fitness function (Eq.~\ref{eq:fitness})\
were\ 
systematically calibrated using automated parameter exploration,\
which we cover later.\
The values listed in the table show the parameter combination that achieved\
the highest energy savings with the least number of constraint violations.\






\subsection{Dynamic Controller Analysis}

This part of the analysis aims to compare our pervasive cloud controller\
with the baseline controller by\
visualising individual actions.\
The use case is the scenario with world-wide data centers\
and other parameters we already described\
in Table~\ref{tab:parameters-simulation}.\

In Fig.~\ref{fig:results-dynamic} we see 6x4 graphs, where the columns
represent data centers. First two rows show geotemporal inputs --\
electricity price and temperatures.\
The next two row shows the dynamic number of active \gls{vm}s\
at the corresponding data center.\
The third row shows the behaviour of the baseline controller\
and the fourth row of the pervasive cloud controller.\
The x-axes of all the graphs cover the same time span and the graphs\
in the same column are aligned to the same x-axis, shown in the bottom.\
Similarly, the graphs in the same row share the same y-axis.\

The electricity prices are lowest in the USA\
(with an increase towards the end of the day),\
followed by Asian locations and the European locations have significantly\
higher prices.\
Temperatures start the lowest in Europe, but then approach 20 C.\
The other locations oscillate around 20 C, except for the peaks in Asian\
locations where 30 C are approached.\
We can see that the baseline method roughly uniformly distributes\
the \gls{vm}s across all the available data centers,\
disregarding the geotemporal inputs.\
The pervasive cloud controller allocates \gls{vm}s in the first 18 hours\
filling out the US capacities, targeting lower electricity prices.\
No \gls{vm}s are allocated in the European locations during this period,\
due to high electricity prices and enough capacity at other locations.\
The Asian locations are initially empty,\
but after the temperature peak is over,\
\gls{vm}s are migrated to the Singapore data center\
and at the end of the first day\
(when Asian electricity prices start to decrease even bellow the US values),\
the Taiwan data center as well.\
This shows us the desired behaviour of the pervasive cloud controller\
where geotemporal inputs are monitored and adapted to by reallocating\
load to the most cost-efficient data center location.\

\subsection{Aggregated Simulation Results}

To give an estimation of the benefits of using our pervasive cloud controller\
in a large-scale scenario described in Table~\ref{tab:parameters-simulation}\
and analyse various environmental parameters,\
we collected the performed actions during the whole simulation and calculated the\
aggregated energy consumption and costs.\ 




\subsubsection{Cost Savings}

\begin{figure}
\vspace{\figtopmargin}
\vspace{-0.3cm}
\includegraphics[width=1.0\columnwidth]{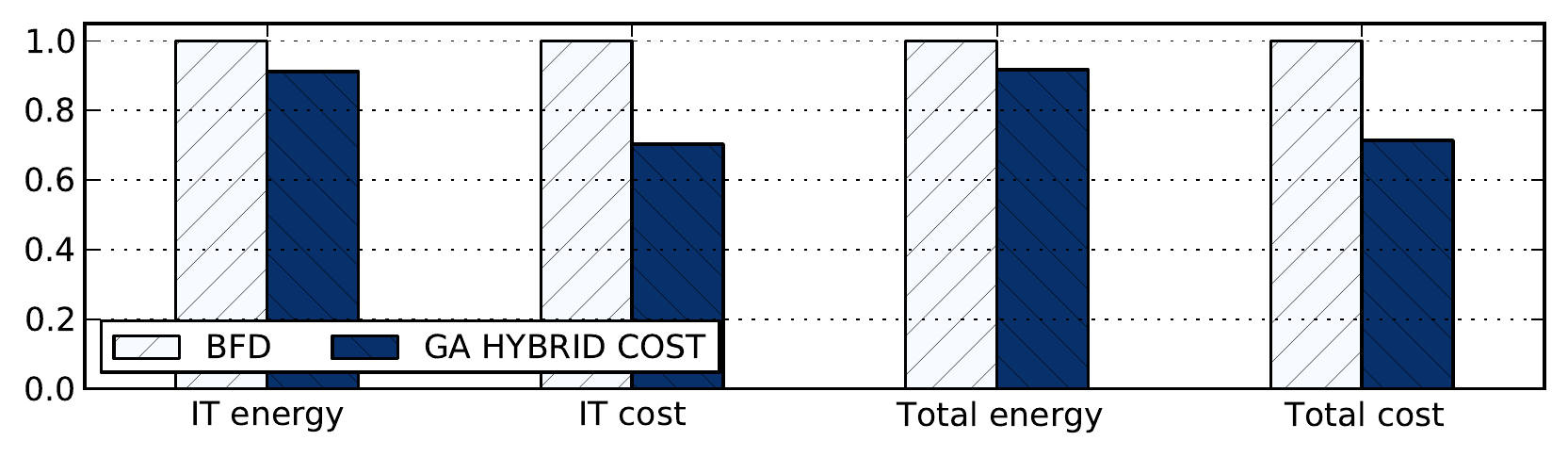}
\vspace{\figcaptionmargin}
\vspace{-0.15cm}
\caption{Normalised energy and costs of the pervasive cloud controller\
compared to the baseline method in a simulation of \vmnumsimulation{} \gls{vm}s.}
\label{fig:results-decisions}
\vspace{\figbottommargin}
\vspace{0.1cm}
\end{figure}

The normalised results of the simulation based on the world-wide dataset\
with \vmnumsimulation{} \gls{vm}s\
are shown in Fig.~\ref{fig:results-decisions}.\
A group of columns is shown for each of the examined quality metrics -- IT energy,\
IT cost, total energy and total cost.\
Inside each group, there is a column for both of the scheduling algorithms:\
the baseline algorithm (BFD)\
and the pervasive cloud controller (GA HYBRID COST).\
The values are normalised as a relative value of\
the baseline algorithm's results.\
The absolute values are listed in Table~\ref{tab:results-decisions}.\
The pervasive cloud controller achieves savings\
of \ensavingsnogeotemp{} in total energy cost compared to the baseline.\
We can see that significant savings can be achieved\
using our pervasive cloud controller,\
which is especially relevant for large cloud providers\
such as Google or Microsoft that spend over \$40M annually on data center\
electricity costs \cite{qureshi_cutting_2009}.\

\begin{table}
\centering
\caption{Absolute energy consumption and costs}
\vspace{\tablecaptionmargin}
\label{tab:results-decisions} 
\begin{tabular}{lrr}
\toprule
{} &        BFD &  GA HYBRID COST \\
\midrule
IT energy (kWh)    &    6226.00 &         5673.63 \\
IT cost (\$)        &     309.40 &          217.64 \\
Total energy (kWh) &    7488.01 &         6869.55 \\
Total cost (\$)     &     370.20 &          264.39 \\
\bottomrule
\end{tabular}
\vspace{-0.5cm}
\end{table}



\subsubsection{Decision Support Component Variation}

\newcommand{\ganesavings}{13\%}
\newcommand{\gantsavings}{7.5\%}

To validate the controller's extensibility and show that it can work\
with different decision support components, we performed\
the same \vmnumsimulation{} \gls{vm} simulation\
with different subsets of the decision support components\
considered by the optimisation engine.\
This analysis also gives an overview of the impact individual geotemporal\
inputs have in the total achieved energy savings.

\begin{figure}
\includegraphics[width=1.0\columnwidth]{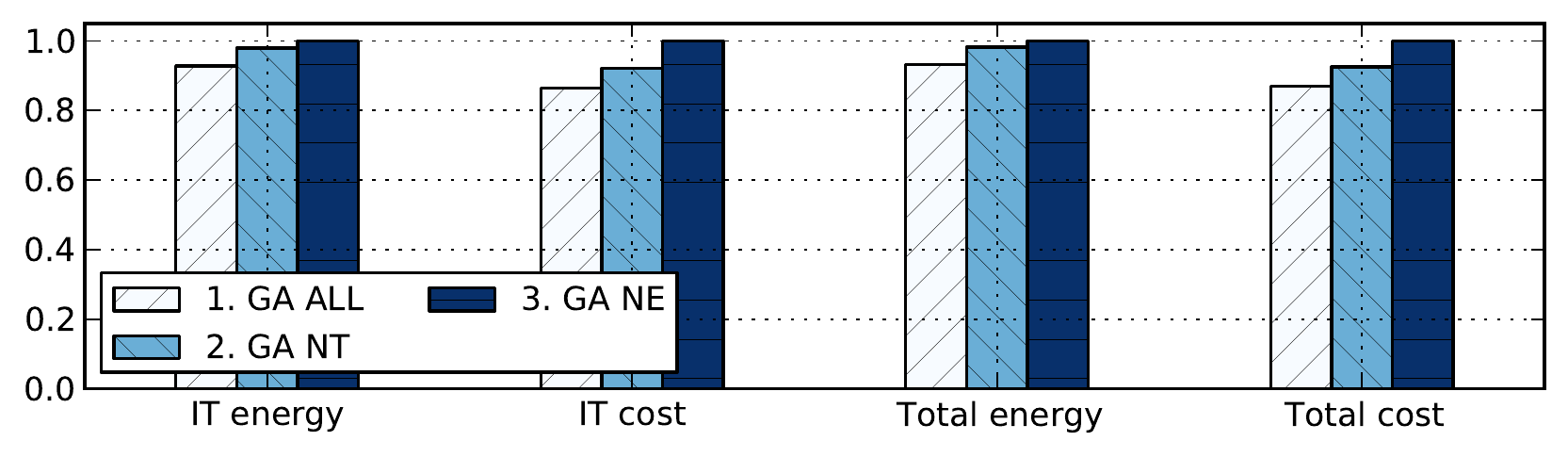}
\vspace{\figcaptionmargin}
\vspace{-0.15cm}
\caption{Normalised energy and costs of the pervasive cloud controller\ 
with various decision support components:\ 
temperature and electricity price (1), no temperature (2)\
and no temperature or electricity cost (3).}
\label{fig:extensibility}
\vspace{-0.4cm}
\end{figure}

The results are shown in Fig.~\ref{fig:extensibility}.\
Each column stands for one of the simulation scenarios\
-- both temperatures and electricity price components (GA ALL),\
electricity price component, but no temperature (GA NT) and\
no electricity price or temperature components (GA NE).\
Total cost savings of \gantsavings{} are achieved when both\
components are considered compared to not considering temperatures.\
Savings are \ganesavings{} when compared to not considering both components,\
which is a significant difference.\
In the same manner we turn on or off certain decision support components\
as a configuration option in the controller's implementation,\
new geotemporal inputs and rules can be added in the future when necessary.\

\subsubsection{Geography Variation} 

Different cloud providers will have different data center locations and\
geographical distributions.\
To estimate the impact of this geographical distribution of data centers\
on the possible cost savings achievable using our scheduling method,\
we simulate its effects on two such scenarios -- the world-wide scenario\
and the USA scenario,\
as described in Section~\ref{sec:evaluation_methodology}.\
Furthermore, as the USA dataset of geotemporal inputs consists of real\
historical electricity price traces (which we did not have to artificially\
adapt to different time zones and local averages) it further testifies to the\
validity of our approach.\
Lastly, even though current cloud providers have incentives to spread\
their data centers further apart to bring services closer\
to a world-wide user base, with the advent of smart buildings\
\cite{privat_smart_2013}\
we might see more localised data center distributions based on\
neighborhood, city or region organisations.

\begin{figure}
\includegraphics[width=1.0\columnwidth]{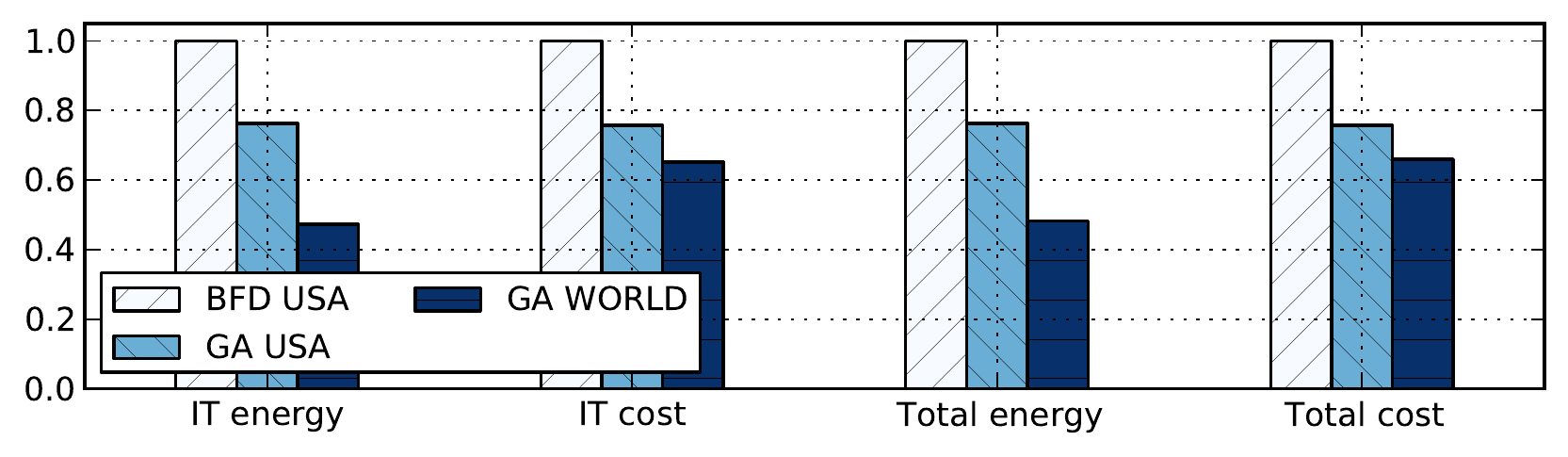}
\vspace{\figcaptionmargin}
\vspace{-0.15cm}
\caption{Normalised energy and costs of the pervasive cloud controller\
compared to the baseline method for the USA and world-wide datasets.}
\label{fig:usa-world}
\vspace{\figbottommargin}
\end{figure}

The results of the simulation for the USA and world-wide dataset\
for \vmnumsimulation{} \gls{vm}s are shown in Fig.~\ref{fig:usa-world}.\
The simulation settings were the same\
we explained in Section~\ref{sec:evaluation_methodology},\
except for the locations of the physical machines.\
The baseline controller was run for the USA dataset (BFD USA),\
and we can see the normalised results compared to this baseline\
for the pervasive cloud controller simulated on the USA dataset (GA USA)\
and the world-wide dataset (GA WORLD).\
It can be seen that  significant cost savings of 24\% are achieved\
even for the USA-only scenario.\
The pervasive cloud controller's energy consumption and costs\
are lower in the world-wide scenario than for the US-only data centers,\
though -- a further 10\% decrease is possible.\



\subsubsection{\gls{qos} Analysis}

\newcommand{\onemigrationpercentage}{$20\%$}
\newcommand{\aggregatedmigrationsmin}{one}
\newcommand{\aggregatedmigrationslow}{two}
\newcommand{\aggregatedmigrationsmax}{three}
\newcommand{\aggregatedmigrationsci}{1.26--1.5}

\begin{figure}
\vspace{\figtopmargin}
\includegraphics[width=1.0\columnwidth]{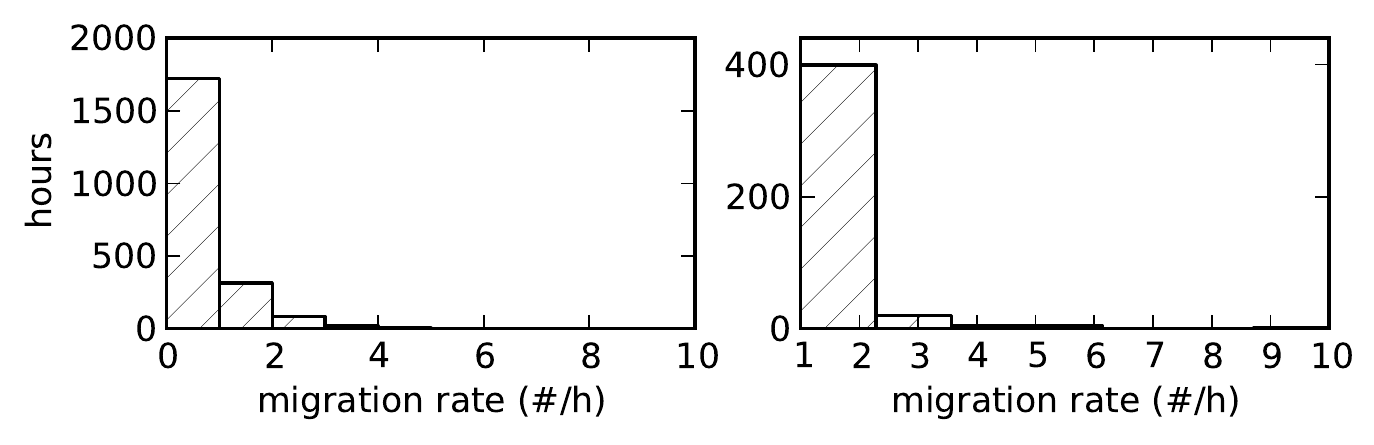}
\vspace{\figcaptionmargin}
\vspace{-0.15cm}
\caption{Hourly migration rate histogram (two zoom levels).}
\label{fig:migrations-rate}
\vspace{-0.2cm}
\end{figure}

Aside from understanding the cost savings\
from the cloud provider's perspective,\
we also have to analyse the \gls{qos}, i.e.\
how the controller affects end users of \gls{vm}s.\
To measure this, we count how often the migration actions\
occur, i.e. the migration rate.\
To get more data,\
we ran the simulation to cover three months.\
A histogram of hourly migration rates of all the \gls{vm}s\
obtained from the simulation of the pervasive cloud controller\
can be seen in Fig.~\ref{fig:migrations-rate}.\
The two plots show different zoom levels, as there are progressively less\ 
hours with higher migration rates.\
Most of the time, no migrations are scheduled,\
with one migration per hour happening\
about \onemigrationpercentage{} of the time.\


\begin{figure}
\vspace{\figtopmargin}
\includegraphics[width=1.0\columnwidth]{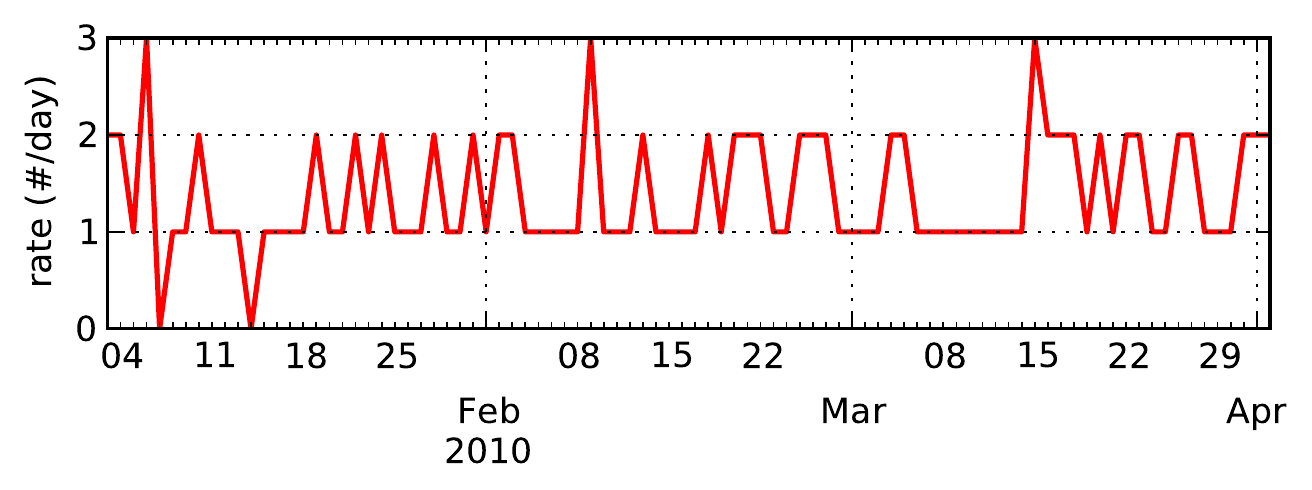}
\vspace{\figcaptionmargin}
\vspace{-0.15cm}
\caption{Aggregated worst-case per-\gls{vm} migration rate.}
\label{fig:migrations-aggregated}
\vspace{\figbottommargin}
\end{figure}

To get a \gls{qos} metric meaningful to the user,\
we group migrations per \gls{vm}\
(as users are only interested in migrations of\
their own \gls{vm}s)\
and process them in a\
function that aggregates migrations over a daily interval.\
We then define the \textit{aggregated worst-case} metric by\ 
counting the migrations per \gls{vm} per day and selecting\
the highest migration count among all the \gls{vm}s in every interval.\
Such a metric could be useful e.g. in defining the lower bound for the\
availability rate in an \gls{sla}.\
The aggregated worst-case migration rate for the simulation\
of the pervasive cloud controller\
is shown in Fig.~\ref{fig:migrations-aggregated}.\
There are \aggregatedmigrationsmin{} or \aggregatedmigrationslow{} migrations\
per \gls{vm} per day most of the time,\
with an occasional case with a higher rate such as the peaks with\
\aggregatedmigrationsmax{} migrations.\

Given that this data is highly dependent of the\
scheduling algorithm parameters used\
and the actual environmental parameters for a specific cloud deployment,\
fitting one specific statistical distribution to the data to get the desired\
percentile value that can be guaranteed in an \gls{sla} would be\
hard to generalise for different use cases and might require manual modelling.\
Instead, we propose applying the\
distribution-independent\
bootstrap confidence interval method \cite{efron_introduction_1994}\
to estimate the aggregated migration rate.\ 
In our simulation,\ 
the 95\% confidence interval for the mean daily per-\gls{vm} migration rate is\ 
\aggregatedmigrationsci{} migrations per day.\


\subsubsection{Genetic Algorithm Parameter Exploration}

To explore how the optimisation engine behaves under different \gls{ga}\
parameters, we ran the simulation with different parameter values and\
compared the resulting energy costs.\
We explored the weights of the different\
fitness function components in (Eq.~\ref{eq:fitness}).\
We developed a method for automatically running the simulation with\
different parameter combinations in the Philharmonic simulator.\ 
We covered a set $\{0, 0.1, 0.2, \ldots\ 1.\}$\
for each of the four weight parameters,\
exploring all the combinations with a constraint that their sum equals 1\
(as only weight ratios make a difference in the \gls{ga},\
not their absolute values),\
resulting in 285 combinations.\
This method can be used to calibrate the controller for different environments\
by finding the parameter combination that achieves\
the highest energy savings or the best \gls{qos},\
similar to how different objectives are optimised\
in a Pareto frontier.\

\begin{figure}
\includegraphics[width=0.9\columnwidth]{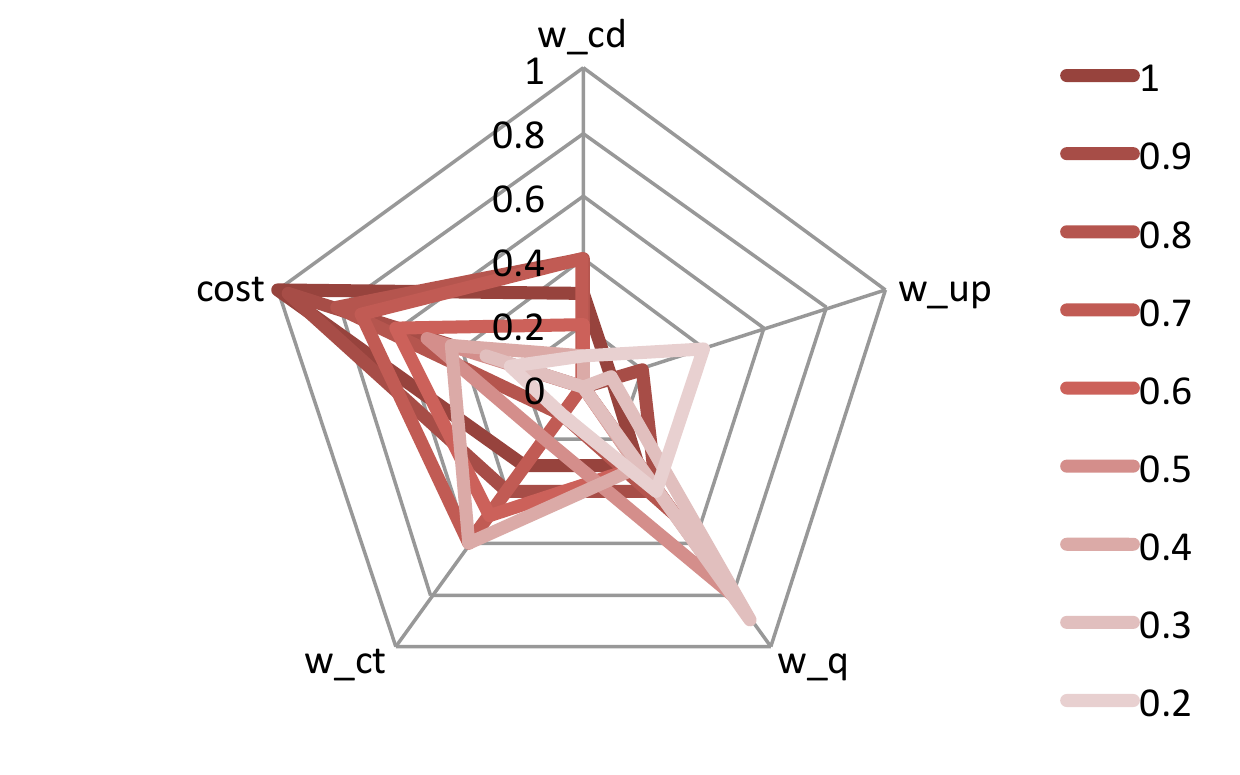}
\vspace{\figcaptionmargin}
\caption{Relative energy costs for different \gls{ga} parameters.}
\label{fig:parameter-exploration}
\end{figure}

A radio chart with the results\
is shown in Fig.~\ref{fig:parameter-exploration}.\
The five axes show the four fitness component weights with the fifth axis\
as the energy cost expressed\
relative to the worst-case combination. One combination is shown as a\
pentagon of the same colour, indicating the 5-tuple\
$(w_{ct}, w_q, w_{cd}, w_{up}, cost)$.\
Combination colour is sorted by relative cost as well, with darker\
colours having higher and lighter colours lower energy costs.
For clarity, we only show a subset of the combinations with the\
rounded relative cost closest to a step of 0.1.\
We can see that the lowest energy costs are achieved for\
high $w_{up}$ and $w_q$ weights, meaning that energy cost\
and a low number of migrations is prioritized.\
Higher energy costs were obtained when constraint satisfaction ($w_{ct}$)\
and \gls{vm} consolidation ($w_{cd}$) is prioritized,\
as neither of these components includes geotemporal inputs.

\subsubsection{Temperature Range Variation} 


\newcommand{\temperaturerangersquared}{0.31}

As different cloud providers have data centers at various locations,\
where temperature ranges can be very different, we analyse the impact of\
temperature range variation on pervasive cloud control effectiveness.\
Temperature variation is affected by the time of the year and the range\
of daily temperatures will vary over time.\
For this reason, we performed the simulation with different starting times\
throughout the year, which resulted in different temperature ranges\
for different simulation runs.

\begin{figure}
\includegraphics[width=1.0\columnwidth]{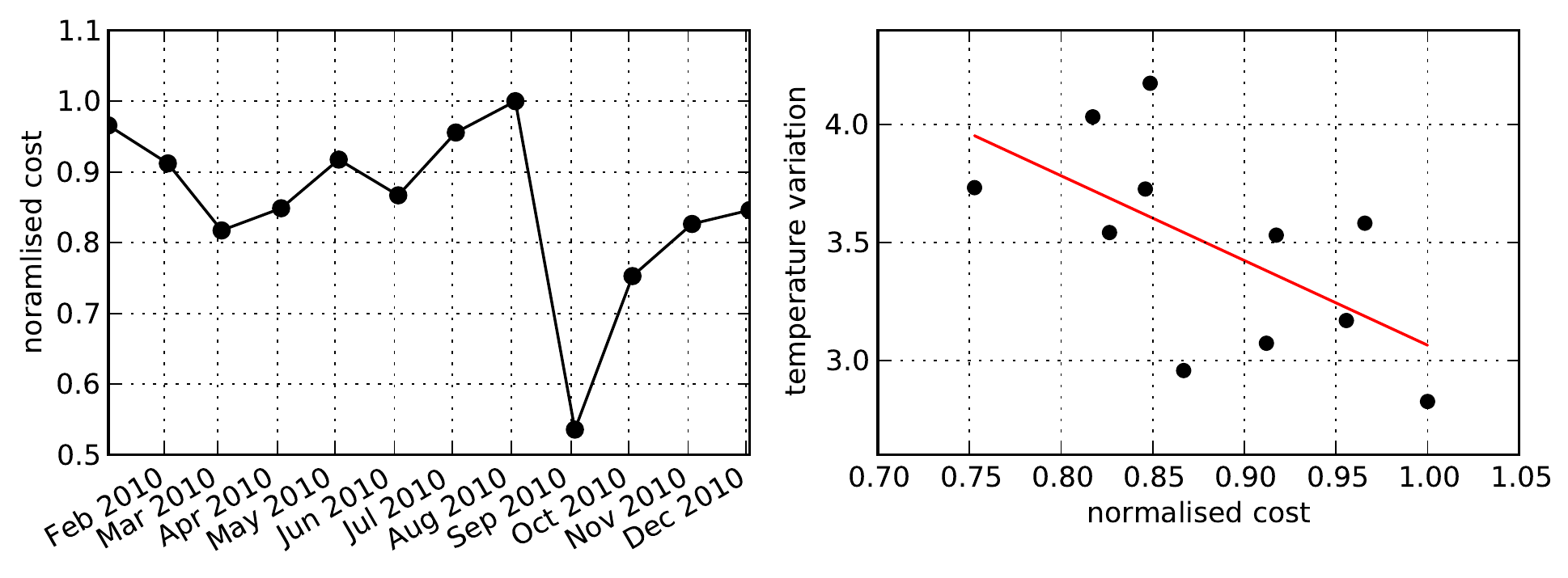}
\vspace{\figcaptionmargin}
\vspace{-0.15cm}
\caption{Energy costs for simulation runs during different months\
of the year (left) and a scatter plot of the same data and the\
temperature variation metric showing linear dependence (right).}
\label{fig:temperature-variation}
\end{figure}

The resulting graphs are shown in Fig.~\ref{fig:temperature-variation}.\
The figure to the left shows the energy cost (normalised as relative\
to the maximum vaule) resulting from the simulation of the pervasive\
cloud controller handling the same \gls{vm} requests, only shifted\
to a different month of the year.\
We can see a gradual trend, with a single sudden drop in October.\
To extract the statistical environment changes between these runs,\
we plotted the same normalised cost as a scatter plot against\
the temperature variation in the figure to the right.\
The temperature variation is calculated as the mean\
of the standard deviations of temperature values\
for individual data center locations.\
Once ordered this way, the trend of the data becomes clearer\
and we calculated a linear correlation between the variables\
with an adjusted $R^2$ of \temperaturerangersquared{} (model shown in red).\
We see that a higher temperature variation\ 
results in lower energy costs.\
This is due to the higher impact avoiding locations with\
unfavourable cooling efficiency conditions has.\

\subsubsection{Data Quality} 

To explore the effect errors in forecasting geotemporal inputs have on\
the pervasive cloud controller's operation, we simulate different\
data quality scenarios.\
%
Additionally, we explore different forecast window sizes\
and their effect on scheduling efficiency.\
A longer forecast window enables the fitness function to\
evaluate the consequences of different management actions\
over a longer interval, reducing the impact of short-term\
geotemporal impact changes, such as electricity price spikes.\

\begin{figure}
\vspace{\figtopmargin}
\includegraphics[width=1.0\columnwidth]{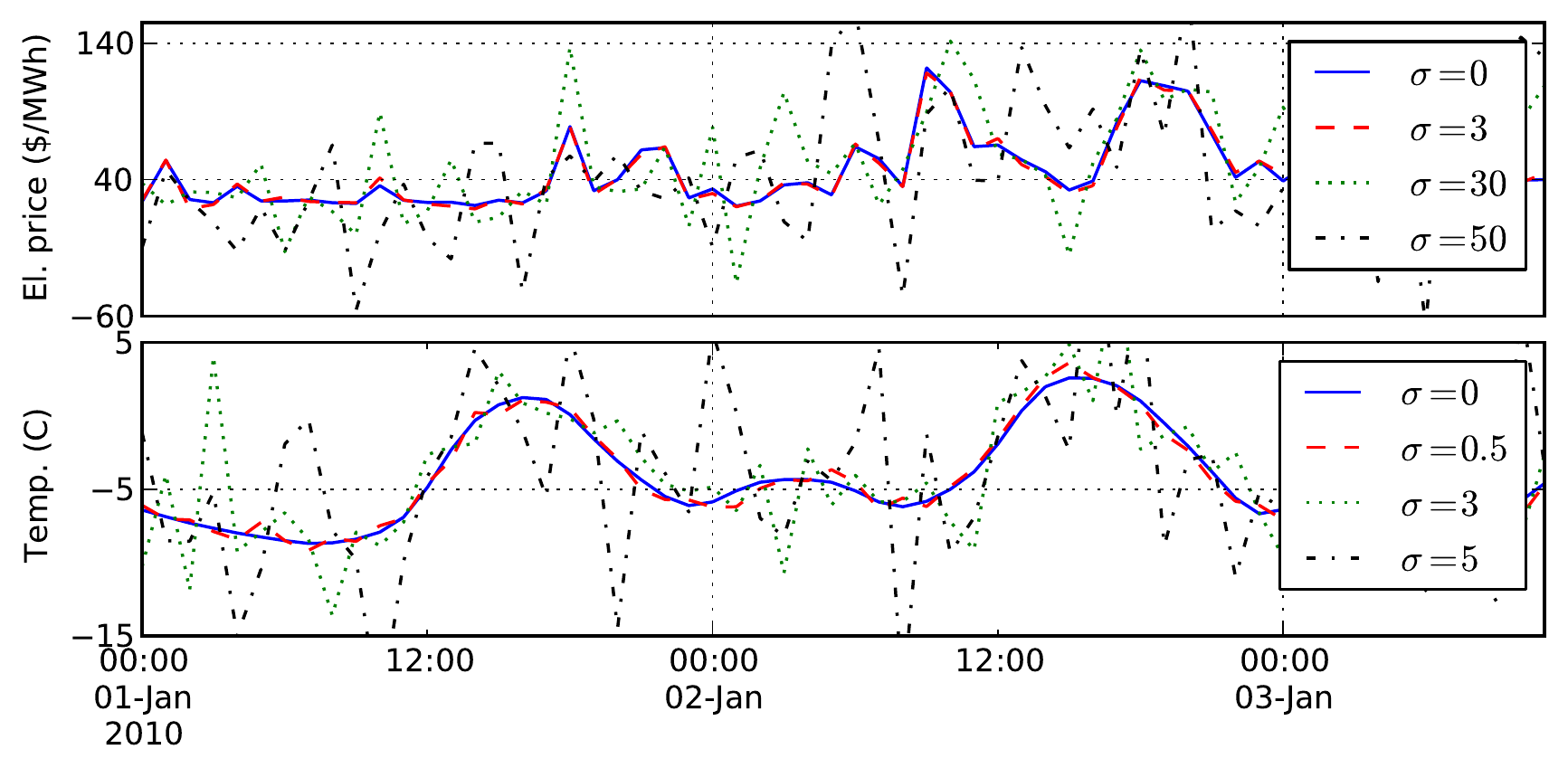}
\vspace{-0.8cm}
\caption{Different levels of forecasting errors applied to\
electricity price and temperature time series (for Mankato, MN).}
\label{fig:forecasting-error-model}
\vspace{-0.6cm}
\end{figure}

The time series provided to the scheduling algorithm\
with different forecasting errors were obtained\
using Eq.~\ref{eq:forecasting_error}\
by selecting different standard deviation ($\sigma_{pred}$) parameters.\
A $\sigma_{pred}$ close to zero represents very accurate forecasting,\
while a higher $\sigma_{pred}$ causes higher signal volatility and\
forecasting errors.\
A segment of the generated time series for one of the cities is\
shown in Fig.~\ref{fig:forecasting-error-model}.\
Both time series are aligned to the same x-axis.\
Each curve represents one $\sigma_{pred}$ scenario.\
It can be seen that smaller error levels\
($\sigma_{pred}$ of 3 \$/MWh and 0.5 C\
for electricity or temperature, respectively)\
still retain the general trend with identifiable peaks and lows.\
Higher error levels\
($\sigma_{pred}$ from 30 to 50 \$/MWh or 3 to 5 C\
for electricity or temperature, respectively)\
start to significantly diverge from the original time series,\
in that peaks are predicted where in reality lows occur and vice versa.\
We simulated forecast window sizes of 4, 12, 24 and 48 hours.\


\begin{figure}
\centering
\vspace{\figtopmargin}
\includegraphics[width=0.8\columnwidth,trim={1.6cm 0 0 0.3cm},clip]{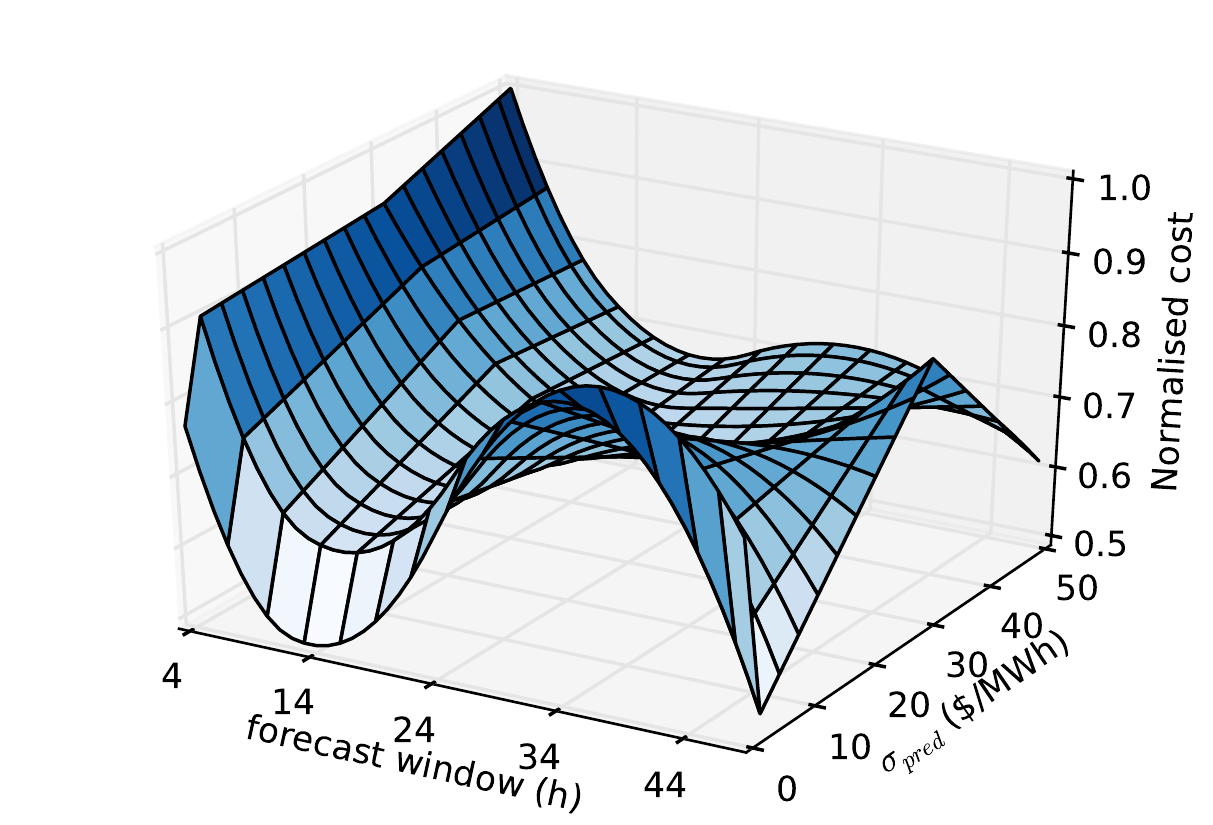}
\vspace{-0.3cm}
\caption{Normalised energy costs resulting from different forecasting errors and forecast window sizes.}
\label{fig:results-forecasting-parameters}
\vspace{\figbottommargin}
\end{figure}

The generated time series with forecasting errors were provided\
to the pervasive cloud controller to base its decisions upon.\
The forecast parameter exploration results are shown in\
Fig.~\ref{fig:results-forecasting-parameters}.\
The 3-dimensional visualisation shows the space\
of forecast window sizes and electricity price $\sigma_{pred}$ values\
on the bottom plane. The z-position on the surface (its height) shows\
the total energy cost (including the migration and cooling overhead),\
normalised relative to the worst case ($fw=4\ h,\ \sigma_{pred}=50\ \$/MWh$).\
Missing data points\ 
were interpolated.\ 
The graph only shows $\sigma_{pred}$ used to model electricity price\
forecasting errors, but a matching $\sigma_{pred}$ for temperature\
from Fig.~\ref{fig:forecasting-error-model} was also applied.\

Looking at the forecast window sizes\
in Fig.~\ref{fig:results-forecasting-parameters},\
we can see that initially bigger windows result in lower costs.\
This trend can clearly be seen from 4 h to 20 h for all $\sigma_{pred}$.\
The trend changes, however, for 24 h and bigger windows.\
Higher or lower savings are visible\ 
and the pattern is more randomised.\
The reason is that both electricity prices and temperatures exhibit a daily\
seasonality effect and extending the window further than 24 h does not\
provide much more information to the controller,\
but increases the problem search space.\
We conclude that the rift-like surface shape in\
the forecast window range from 12 to 24 hours\
represents an optimal size for the given geotemporal inputs.

The forecasting error dimension shows\
deviations of around 25\% between large forecasting errors and\
perfect knowledge.\
This can be attributed to the fact that large forecasting errors\
mislead the controller into placing \gls{vm}s in areas where geotemporal\
inputs are in fact worse, so both energy cost losses\
and migration overheads are incurred.\ 
Smaller forecasting errors result in lower energy costs,\
which shows the importance of having accurate forecasting methods\
(or data sources) when managing clouds based on geotemporal inputs.\
Based on our simulation, $\sigma_{pred}$ of 3 \$/MWh\
(\gls{mse} of $\approx 9$) and 0.5 C\
(\gls{mse} of $\approx 0.25$) or less is\
necessary for feasible cost savings.\

\subsection{Cloud Provider Guidelines Case Study}

To show the usage potential of the collected measurements as\ 
guidelines, we performed a case study for several different cloud providers.\
The results are shown in Table~\ref{tab:case_study}.\
The annual electricity cost estimations\
are obtained from \cite{qureshi_cutting_2009}.\
We selected cloud providers of different scale\
(e.g. A and C).\
We compare several environment conditions,\
namely the temperature variation $temp\_var$ and\
data quality metrics $fw$ and $\sigma_{pred}$,\
considering different combinations as hypothetical scenarios that\
cloud providers might be in.\
The cost factor is calculated based on the already analysed\
temperature variation linear model\
from Fig.~\ref{fig:temperature-variation} and\
the forecast data quality impact results\
from Fig.~\ref{fig:results-forecasting-parameters}.\
Finally, we show the order of magnitude of the cost savings\ 
based on the cost factor.\

\vspace{\tabletopmargin}
\begin{table}[H]
\centering
\caption{Cloud provider guidelines case study.}
\vspace{\tablecaptionmargin}
\label{tab:case_study} 
\begin{tabular}{llrrrrl}
\toprule
provider & electricity &  $fw$ &  $\sigma_{pred}$ &  $temp\_var$ &  factor & savings\\
\midrule
        A &             \$38M &  14 &     10 &       3.5 &        0.565 &         \$16.5M \\
     B &             \$36M &  48 &     30 &       2.5 &        0.989 &          \$0.4M \\
     C &             \$12M &  24 &     10 &       4.0 &        0.536 &          \$5.6M \\
\bottomrule
\end{tabular}
\end{table}
\vspace{\tablebottommargin}

The results show that significant savings are possible\
using pervasive cloud control with appropriate environmental conditions.\
The B scenario, however, shows that even with high initial costs,\
having bad forecast data quality and a low temperature variation\
can result in lower savings,\
perhaps not enough of an incentive to apply our method.\
On the other hand, even a smaller provider, such as C,\
can achieve promising savings\
in a favourable environment.\ 

\vspace{-0.4cm}
\section{Conclusion}


In this paper we presented an approach for pervasive cloud control\
under geotemporal inputs, such as real-time electricity pricing\
and temperature-dependent cooling efficiency.\
The solution is designed for extensibility\
with new geotemporal inputs and cloud regulation mechanisms\
through a modular\
decision support component system and\
a forward-compatible optimisation engine.\
We presented a proof-of-concept controller implementation\
combining forecast-based planning and\
a hybrid genetic algorithm\
with greedy local optimisation.\
The genetic algorithm approach was extended\
with partial population propagation.\

The approach was evaluated in a simulation based on real\
traces of temperatures and electricity prices.\
We estimated energy cost savings\
of up to \ensavingsnogeotemp{}\
compared to a baseline cloud control method\
that applies \gls{vm} consolidation without considering geotemporal inputs.\
We analysed per-\gls{vm} migrations\ 
to show that no significant\
\gls{qos} impact is incurred in the process.\
We evaluated different parameters such as geographical data center distributions and\
forecast data quality\ 
as cloud provider guidelines\
to find conditions fit for pervasive cloud control.\

The questions that remain open are how to extend the method\
for integrated forecasting of arbitrary time series data,\
e.g. application-level\
load predictions or local renewable energy availability.\
Additionally, it would be beneficial to research the method\
in the context of\ 
containers and stateless applications which enable much more\
efficient computation migration.\
We plan to research these topics in our future work.



%

%

\ifCLASSOPTIONcompsoc
  \section*{Acknowledgments}
\else
  \section*{Acknowledgment}
\fi

The work described in this paper has been funded\
through the Haley project (Holistic Energy Efficient Hybrid Clouds)\
as part of the TU Vienna Distinguished Young Scientist Award 2011.

\ifCLASSOPTIONcaptionsoff
  \newpage
\fi



\bibliographystyle{IEEEtran-kermit}
\bibliography{ms.bib}

\begin{thebibliography}{10}
\providecommand{\url}[1]{#1}
\csname url@samestyle\endcsname
\providecommand{\newblock}{\relax}
\providecommand{\bibinfo}[2]{#2}
\providecommand{\BIBentrySTDinterwordspacing}{\spaceskip=0pt\relax}
\providecommand{\BIBentryALTinterwordstretchfactor}{4}
\providecommand{\BIBentryALTinterwordspacing}{\spaceskip=\fontdimen2\font plus
\BIBentryALTinterwordstretchfactor\fontdimen3\font minus
  \fontdimen4\font\relax}
\providecommand{\BIBforeignlanguage}[2]{{%
\expandafter\ifx\csname l@#1\endcsname\relax
\typeout{** WARNING: IEEEtran.bst: No hyphenation pattern has been}%
\typeout{** loaded for the language `#1'. Using the pattern for}%
\typeout{** the default language instead.}%
\else
\language=\csname l@#1\endcsname
\fi
#2}}
\providecommand{\BIBdecl}{\relax}
\BIBdecl

\bibitem{doyle_stratus:_2013}
J.~Doyle, R.~Shorten, and D.~O'Mahony, ``Stratus: {Load} {Balancing} the
  {Cloud} for {Carbon} {Emissions} {Control},'' \emph{IEEE Transactions on
  Cloud Computing}, vol.~1, no.~1, pp. 1--1, Jan. 2013.

\bibitem{kesavan_practical_2013}
M.~Kesavan, I.~Ahmad, O.~Krieger, R.~Soundararajan, A.~Gavrilovska, and
  K.~Schwan, ``Practical {Compute} {Capacity} {Management} for {Virtualized}
  {Datacenters},'' \emph{IEEE Transactions on Cloud Computing}, vol.~1, no.~1,
  pp. 1--1, Jan. 2013.

\bibitem{mastroianni_probabilistic_2013}
C.~Mastroianni, M.~Meo, and G.~Papuzzo, ``Probabilistic {Consolidation} of
  {Virtual} {Machines} in {Self}-{Organizing} {Cloud} {Data} {Centers},''
  \emph{IEEE Transactions on Cloud Computing}, vol.~1, no.~2, pp. 215--228,
  Jul. 2013.

\bibitem{xu_temperature_2013}
\BIBentryALTinterwordspacing
H.~Xu, C.~Feng, and B.~Li, ``Temperature aware workload management in
  geo-distributed datacenters,'' in \emph{Proceedings of the {ACM}
  {SIGMETRICS}/international conference on {Measurement} and modeling of
  computer systems}, vol.~41.\hskip 1em plus 0.5em minus 0.4em\relax ACM, 2013,
  pp. 373--374.
\BIBentrySTDinterwordspacing

\bibitem{christy_sujatha_energy_2011}
D.~Christy~Sujatha and S.~Abimannan, ``Energy {Efficient} {Free} {Cooling}
  {System} for {Data} {Centers},'' in \emph{2011 {IEEE} {Third} {International}
  {Conference} on {Cloud} {Computing} {Technology} and {Science} ({CloudCom})},
  2011, pp. 646--651.

\bibitem{_digitalocean_????}
\BIBentryALTinterwordspacing
``{DigitalOcean}.'' [Online]. Available: \url{https://www.digitalocean.com/}
\BIBentrySTDinterwordspacing

\bibitem{_openstack_????}
\BIBentryALTinterwordspacing
``{OpenStack} {Open} {Source} {Cloud} {Computing} {Software}.'' [Online].
  Available: \url{http://www.openstack.org/}
\BIBentrySTDinterwordspacing

\bibitem{goiri_parasol_2013}
\BIBentryALTinterwordspacing
I.~Goiri, W.~Katsak, K.~Le, T.~D. Nguyen, and R.~Bianchini, ``Parasol and
  greenswitch: {Managing} datacenters powered by renewable energy,'' in
  \emph{{ACM} {SIGARCH} {Computer} {Architecture} {News}}, vol.~41.\hskip 1em
  plus 0.5em minus 0.4em\relax ACM, 2013, pp. 51--64.
\BIBentrySTDinterwordspacing

\bibitem{lucanin_take_2013-1}
\BIBentryALTinterwordspacing
D.~Lučanin and I.~Brandic, ``Take a break: cloud scheduling optimized for
  real-time electricity pricing,'' in \emph{Cloud and {Green} {Computing}
  ({CGC}), 2013 {Third} {International} {Conference} on}.\hskip 1em plus 0.5em
  minus 0.4em\relax IEEE, 2013, pp. 113--120.
\BIBentrySTDinterwordspacing

\bibitem{michalewicz_genetic_1996}
\BIBentryALTinterwordspacing
Z.~Michalewicz, \emph{Genetic algorithms+ data structures= evolution
  programs}.\hskip 1em plus 0.5em minus 0.4em\relax Springer Science \&
  Business Media, 1996.
\BIBentrySTDinterwordspacing

\bibitem{cauwer_study_2013}
M.~D. Cauwer and B.~O'Sullivan, ``A {Study} of {Electricity} {Price} {Features}
  on {Distributed} {Internet} {Data} {Centers},'' in \emph{Economics of
  {Grids}, {Clouds}, {Systems}, and {Services} - 10th {International}
  {Conference}, {GECON} 2013, {Zaragoza}, {Spain}, {September} 18-20, 2013.
  {Proceedings}}, ser. Lecture {Notes} in {Computer} {Science}, J.~Altmann,
  K.~Vanmechelen, and O.~F. Rana, Eds., vol. 8193.\hskip 1em plus 0.5em minus
  0.4em\relax Springer, 2013, pp. 60--73.

\bibitem{_time_2015}
\BIBentryALTinterwordspacing
``\BIBforeignlanguage{en}{Time series},'' Mar. 2015, page Version ID:
  651637275. [Online]. Available:
  \url{http://en.wikipedia.org/w/index.php?title=Time_series&oldid=651637275#Notation}
\BIBentrySTDinterwordspacing

\bibitem{beloglazov_energy-aware_2012}
\BIBentryALTinterwordspacing
A.~Beloglazov, J.~Abawajy, and R.~Buyya, ``Energy-aware resource allocation
  heuristics for efficient management of data centers for {Cloud} computing,''
  \emph{Future Generation Computer Systems}, vol.~28, no.~5, pp. 755--768, May
  2012.
\BIBentrySTDinterwordspacing

\end{thebibliography}


\begin{thebibliography}{10}
\providecommand{\url}[1]{#1}
\csname url@samestyle\endcsname
\providecommand{\newblock}{\relax}
\providecommand{\bibinfo}[2]{#2}
\providecommand{\BIBentrySTDinterwordspacing}{\spaceskip=0pt\relax}
\providecommand{\BIBentryALTinterwordstretchfactor}{4}
\providecommand{\BIBentryALTinterwordspacing}{\spaceskip=\fontdimen2\font plus
\BIBentryALTinterwordstretchfactor\fontdimen3\font minus
  \fontdimen4\font\relax}
\providecommand{\BIBforeignlanguage}[2]{{%
\expandafter\ifx\csname l@#1\endcsname\relax
\typeout{** WARNING: IEEEtran.bst: No hyphenation pattern has been}%
\typeout{** loaded for the language `#1'. Using the pattern for}%
\typeout{** the default language instead.}%
\else
\language=\csname l@#1\endcsname
\fi
#2}}
\providecommand{\BIBdecl}{\relax}
\BIBdecl

\bibitem{jonathan_koomey_growth_2011}
\BIBentryALTinterwordspacing
{Jonathan Koomey}, ``Growth in {Data} center electricity use 2005 to 2010,''
  Analytics Press, Oakland, CA, Tech. Rep., Aug. 2011. [Online]. Available:
  \url{http://www.analyticspress.com/datacenters.html}
\BIBentrySTDinterwordspacing

\bibitem{qureshi_cutting_2009}
\BIBentryALTinterwordspacing
A.~Qureshi, R.~Weber, H.~Balakrishnan, J.~Guttag, and B.~Maggs, ``Cutting the
  electric bill for internet-scale systems,'' \emph{SIGCOMM Comput. Commun.
  Rev.}, vol.~39, no.~4, pp. 123--134, Aug. 2009.
\BIBentrySTDinterwordspacing

\bibitem{privat_smart_2013}
\BIBentryALTinterwordspacing
G.~Privat, ``Smart {Building} {Functional} {Architecture},
  {FI}.{ICT}-2011-285135 {FINSENY} {D}4.3,'' Tech. Rep., 2013. [Online].
  Available:
  \url{http://www.fi-ppp-finseny.eu/wp-content/uploads/2013/04/FINSENY_D4.3_v1.01.pdf}
\BIBentrySTDinterwordspacing

\bibitem{yang_integrating_2013}
\BIBentryALTinterwordspacing
X.~Yang, Z.~Zhou, S.~Wallace, Z.~Lan, W.~Tang, S.~Coghlan, and M.~E. Papka,
  ``Integrating {Dynamic} {Pricing} of {Electricity} into {Energy} {Aware}
  {Scheduling} for {HPC} {Systems},'' in \emph{Proceedings of the
  {International} {Conference} on {High} {Performance} {Computing},
  {Networking}, {Storage} and {Analysis}}, ser. {SC} '13.\hskip 1em plus 0.5em
  minus 0.4em\relax New York, NY, USA: ACM, 2013, pp. 60:1--60:11.
\BIBentrySTDinterwordspacing

\bibitem{weron_modeling_2006}
R.~Weron, \emph{Modeling and {Forecasting} {Electricity} {Loads} and {Prices}:
  {A} {Statistical} {Approach}}, 1st~ed.\hskip 1em plus 0.5em minus 0.4em\relax
  Wiley, Dec. 2006.

\bibitem{barroso_datacenter_2009}
\BIBentryALTinterwordspacing
L.~A. Barroso and U.~Hölzle, ``The datacenter as a computer: {An} introduction
  to the design of warehouse-scale machines,'' \emph{Synthesis lectures on
  computer architecture}, vol.~4, no.~1, pp. 1--108, 2009.
\BIBentrySTDinterwordspacing

\bibitem{goiri_parasol_2013}
\BIBentryALTinterwordspacing
I.~Goiri, W.~Katsak, K.~Le, T.~D. Nguyen, and R.~Bianchini, ``Parasol and
  greenswitch: {Managing} datacenters powered by renewable energy,'' in
  \emph{{ACM} {SIGARCH} {Computer} {Architecture} {News}}, vol.~41.\hskip 1em
  plus 0.5em minus 0.4em\relax ACM, 2013, pp. 51--64.
\BIBentrySTDinterwordspacing

\bibitem{le_reducing_2011}
\BIBentryALTinterwordspacing
K.~Le, R.~Bianchini, J.~Zhang, Y.~Jaluria, J.~Meng, and T.~D. Nguyen,
  ``Reducing {Electricity} {Cost} {Through} {Virtual} {Machine} {Placement} in
  {High} {Performance} {Computing} {Clouds},'' in \emph{Proceedings of 2011
  {International} {Conference} for {High} {Performance} {Computing},
  {Networking}, {Storage} and {Analysis}}, ser. {SC} '11.\hskip 1em plus 0.5em
  minus 0.4em\relax New York, NY, USA: ACM, 2011, pp. 22:1--22:12.
\BIBentrySTDinterwordspacing

\bibitem{liu_data_2013}
\BIBentryALTinterwordspacing
Z.~Liu, A.~Wierman, Y.~Chen, B.~Razon, and N.~Chen, ``Data center demand
  response: {Avoiding} the coincident peak via workload shifting and local
  generation,'' \emph{Performance Evaluation}, vol.~70, no.~10, pp. 770--791,
  Oct. 2013.
\BIBentrySTDinterwordspacing

\bibitem{berl_modelling_2013}
\BIBentryALTinterwordspacing
A.~Berl, G.~Lovász, F.~von Tüllenburg, and H.~de~Meer, ``Modelling {Power}
  {Adaption} {Flexibility} of {Data} {Centres} for {Demand}-{Response}
  {Management},'' in \emph{Energy {Efficiency} in {Large} {Scale} {Distributed}
  {Systems}}.\hskip 1em plus 0.5em minus 0.4em\relax Springer, 2013, pp.
  63--66.
\BIBentrySTDinterwordspacing

\bibitem{xu_temperature_2013}
\BIBentryALTinterwordspacing
H.~Xu, C.~Feng, and B.~Li, ``Temperature aware workload management in
  geo-distributed datacenters,'' in \emph{Proceedings of the {ACM}
  {SIGMETRICS}/international conference on {Measurement} and modeling of
  computer systems}, vol.~41.\hskip 1em plus 0.5em minus 0.4em\relax ACM, 2013,
  pp. 373--374.
\BIBentrySTDinterwordspacing

\bibitem{feller_snooze:_2012}
\BIBentryALTinterwordspacing
E.~Feller, L.~Rilling, and C.~Morin, ``Snooze: {A} {Scalable} and {Autonomic}
  {Virtual} {Machine} {Management} {Framework} for {Private} {Clouds},'' in
  \emph{Proceedings of the 2012 12th {IEEE}/{ACM} {International} {Symposium}
  on {Cluster}, {Cloud} and {Grid} {Computing} ({Ccgrid} 2012)}, ser. {CCGRID}
  '12.\hskip 1em plus 0.5em minus 0.4em\relax Washington, DC, USA: IEEE
  Computer Society, 2012, pp. 482--489.
\BIBentrySTDinterwordspacing

\bibitem{beloglazov_managing_2013}
A.~Beloglazov and R.~Buyya, ``Managing {Overloaded} {Hosts} for {Dynamic}
  {Consolidation} of {Virtual} {Machines} in {Cloud} {Data} {Centers} {Under}
  {Quality} of {Service} {Constraints},'' \emph{IEEE Transactions on Parallel
  and Distributed Systems}, vol.~24, no.~7, pp. 1366--1379, 2013.

\bibitem{maurer_enacting_2011}
M.~Maurer, I.~Brandic, and R.~Sakellariou, ``Enacting {SLAs} in clouds using
  rules,'' \emph{Euro-Par 2011 Parallel Processing}, pp. 455--466, 2011.

\bibitem{liu_performance_2011}
\BIBentryALTinterwordspacing
H.~Liu, C.-Z. Xu, H.~Jin, J.~Gong, and X.~Liao, ``Performance and energy
  modeling for live migration of virtual machines,'' in \emph{Proceedings of
  the 20th international symposium on {High} performance distributed
  computing}, 2011, pp. 171--182.
\BIBentrySTDinterwordspacing

\bibitem{beloglazov_openstack_2014}
\BIBentryALTinterwordspacing
A.~Beloglazov and R.~Buyya, ``{OpenStack} {Neat}: {A} {Framework} for {Dynamic}
  and {Energy}-{Efficient} {Consolidation} of {Virtual} {Machines} in
  {OpenStack} {Clouds},'' \emph{Concurrency and Computation: Practice and
  Experience (CCPE)}, 2014.
\BIBentrySTDinterwordspacing

\bibitem{alfeld_toward_2012}
\BIBentryALTinterwordspacing
S.~Alfeld, C.~Barford, and P.~Barford, ``Toward an analytic framework for the
  electrical power grid,'' in \emph{Proceedings of the 3rd {International}
  {Conference} on {Future} {Energy} {Systems}: {Where} {Energy}, {Computing}
  and {Communication} {Meet}}, ser. e-{Energy} '12.\hskip 1em plus 0.5em minus
  0.4em\relax New York, NY, USA: ACM, 2012, pp. 9:1--9:4.
\BIBentrySTDinterwordspacing

\bibitem{_forecast_2015}
\BIBentryALTinterwordspacing
``Forecast,'' 2015. [Online]. Available: \url{http://forecast.io/}
\BIBentrySTDinterwordspacing

\bibitem{rao_minimizing_2010}
L.~Rao, X.~Liu, L.~Xie, and W.~Liu, ``Minimizing {Electricity} {Cost}:
  {Optimization} of {Distributed} {Internet} {Data} {Centers} in a
  {Multi}-{Electricity}-{Market} {Environment},'' in \emph{2010 {Proceedings}
  {IEEE} {INFOCOM}}, Mar. 2010, pp. 1--9.

\bibitem{lin_online_2012}
M.~Lin, Z.~Liu, A.~Wierman, and L.~Andrew, ``Online algorithms for geographical
  load balancing,'' in \emph{Green {Computing} {Conference} ({IGCC}), 2012
  {International}}, Jun. 2012, pp. 1--10.

\bibitem{li_towards_2012}
J.~Li, Z.~Li, K.~Ren, and X.~Liu, ``Towards {Optimal} {Electric} {Demand}
  {Management} for {Internet} {Data} {Centers},'' \emph{IEEE Transactions on
  Smart Grid}, vol.~3, no.~1, pp. 183--192, Mar. 2012.

\bibitem{doyle_stratus:_2013}
J.~Doyle, R.~Shorten, and D.~O'Mahony, ``Stratus: {Load} {Balancing} the
  {Cloud} for {Carbon} {Emissions} {Control},'' \emph{IEEE Transactions on
  Cloud Computing}, vol.~1, no.~1, pp. 1--1, Jan. 2013.

\bibitem{buchbinder_online_2011}
\BIBentryALTinterwordspacing
N.~Buchbinder, N.~Jain, and I.~Menache, ``Online {Job}-{Migration} for
  {Reducing} the {Electricity} {Bill} in the {Cloud},'' in \emph{{NETWORKING}
  2011}, ser. Lecture {Notes} in {Computer} {Science}, J.~Domingo-Pascual,
  P.~Manzoni, S.~Palazzo, A.~Pont, and C.~Scoglio, Eds.\hskip 1em plus 0.5em
  minus 0.4em\relax Springer Berlin Heidelberg, Jan. 2011, no. 6640, pp.
  172--185.
\BIBentrySTDinterwordspacing

\bibitem{guler_cutting_2013}
H.~Guler, B.~Cambazoglu, and O.~Ozkasap, ``Cutting {Down} the {Energy} {Cost}
  of {Geographically} {Distributed} {Cloud} {Data} {Centers},'' in \emph{Energy
  {Efficiency} in {Large} {Scale} {Distributed} {Systems}}.\hskip 1em plus
  0.5em minus 0.4em\relax Vienna: Springer Berlin Heidelberg, 2013, pp.
  279--286.

\bibitem{liu_renewable_2012}
\BIBentryALTinterwordspacing
Z.~Liu, Y.~Chen, C.~Bash, A.~Wierman, D.~Gmach, Z.~Wang, M.~Marwah, and
  C.~Hyser, ``Renewable and cooling aware workload management for sustainable
  data centers,'' in \emph{Proceedings of the 12th {ACM}
  {SIGMETRICS}/{PERFORMANCE} joint international conference on {Measurement}
  and {Modeling} of {Computer} {Systems}}, ser. {SIGMETRICS} '12.\hskip 1em
  plus 0.5em minus 0.4em\relax New York, NY, USA: ACM, 2012, pp. 175--186.
\BIBentrySTDinterwordspacing

\bibitem{ren_provably-efficient_2012}
S.~Ren, Y.~He, and F.~Xu, ``Provably-{Efficient} {Job} {Scheduling} for
  {Energy} and {Fairness} in {Geographically} {Distributed} {Data} {Centers},''
  in \emph{2012 {IEEE} 32nd {International} {Conference} on {Distributed}
  {Computing} {Systems} ({ICDCS})}, Jun. 2012, pp. 22 --31.

\bibitem{kesavan_practical_2013}
M.~Kesavan, I.~Ahmad, O.~Krieger, R.~Soundararajan, A.~Gavrilovska, and
  K.~Schwan, ``Practical {Compute} {Capacity} {Management} for {Virtualized}
  {Datacenters},'' \emph{IEEE Transactions on Cloud Computing}, vol.~1, no.~1,
  pp. 1--1, Jan. 2013.

\bibitem{mastroianni_probabilistic_2013}
C.~Mastroianni, M.~Meo, and G.~Papuzzo, ``Probabilistic {Consolidation} of
  {Virtual} {Machines} in {Self}-{Organizing} {Cloud} {Data} {Centers},''
  \emph{IEEE Transactions on Cloud Computing}, vol.~1, no.~2, pp. 215--228,
  Jul. 2013.

\bibitem{cauwer_study_2013}
M.~D. Cauwer and B.~O'Sullivan, ``A {Study} of {Electricity} {Price} {Features}
  on {Distributed} {Internet} {Data} {Centers},'' in \emph{Economics of
  {Grids}, {Clouds}, {Systems}, and {Services} - 10th {International}
  {Conference}, {GECON} 2013, {Zaragoza}, {Spain}, {September} 18-20, 2013.
  {Proceedings}}, ser. Lecture {Notes} in {Computer} {Science}, J.~Altmann,
  K.~Vanmechelen, and O.~F. Rana, Eds., vol. 8193.\hskip 1em plus 0.5em minus
  0.4em\relax Springer, 2013, pp. 60--73.

\bibitem{abbasi_dynamic_2011}
Z.~Abbasi, T.~Mukherjee, G.~Varsamopoulos, and S.~K.~S. Gupta, ``Dynamic
  hosting management of web based applications over clouds,'' in \emph{2011
  18th {International} {Conference} on {High} {Performance} {Computing}
  ({HiPC})}, Dec. 2011, pp. 1--10.

\bibitem{zhu_real-time_2014}
X.~Zhu, L.~Yang, H.~Chen, J.~Wang, S.~Yin, and X.~Liu, ``Real-{Time} {Tasks}
  {Oriented} {Energy}-{Aware} {Scheduling} in {Virtualized} {Clouds},''
  \emph{IEEE Transactions on Cloud Computing}, vol.~2, no.~2, pp. 168--180,
  Apr. 2014.

\bibitem{hu_multiairport_2007}
X.-B. Hu, W.-H. Chen, and E.~Di~Paolo, ``Multiairport {Capacity} {Management}:
  {Genetic} {Algorithm} {With} {Receding} {Horizon},'' \emph{IEEE Transactions
  on Intelligent Transportation Systems}, vol.~8, no.~2, pp. 254--263, Jun.
  2007.

\bibitem{kolodziej_hierarchical_2013}
\BIBentryALTinterwordspacing
J.~Kolodziej, S.~U. Khan, L.~Wang, A.~Byrski, N.~Min-Allah, and S.~A. Madani,
  ``Hierarchical genetic-based grid scheduling with energy optimization,''
  \emph{Cluster Computing}, vol.~16, no.~3, pp. 591--609, 2013.
\BIBentrySTDinterwordspacing

\bibitem{larumbe_tabu_2013}
F.~Larumbe and B.~Sanso, ``A {Tabu} {Search} {Algorithm} for the {Location} of
  {Data} {Centers} and {Software} {Components} in {Green} {Cloud} {Computing}
  {Networks},'' \emph{IEEE Transactions on Cloud Computing}, vol.~1, no.~1, pp.
  22--35, Jan. 2013.

\bibitem{meisner_powernap:_2009}
\BIBentryALTinterwordspacing
D.~Meisner, B.~T. Gold, and T.~F. Wenisch, ``{PowerNap}: eliminating server
  idle power,'' \emph{SIGPLAN Not.}, vol.~44, no.~3, pp. 205--216, Mar. 2009.
\BIBentrySTDinterwordspacing

\bibitem{goldberg_genetic_1989}
D.~E. Goldberg, \emph{Genetic algorithms in search, optimization, and machine
  learning}.\hskip 1em plus 0.5em minus 0.4em\relax Addison-wesley Reading
  Menlo Park, 1989, vol. 412.

\bibitem{beloglazov_energy-aware_2012}
\BIBentryALTinterwordspacing
A.~Beloglazov, J.~Abawajy, and R.~Buyya, ``Energy-aware resource allocation
  heuristics for efficient management of data centers for {Cloud} computing,''
  \emph{Future Generation Computer Systems}, vol.~28, no.~5, pp. 755--768, May
  2012.
\BIBentrySTDinterwordspacing

\bibitem{melard_automatic_2000}
G.~Mélard and J.-M. Pasteels, ``Automatic {ARIMA} modeling including
  interventions, using time series expert software,'' \emph{International
  Journal of Forecasting}, vol.~16, no.~4, pp. 497--508, Oct. 2000.

\bibitem{drazen_lucanin_philharmonic_2014}
\BIBentryALTinterwordspacing
{Dražen Lučanin}, ``Philharmonic,'' 2014. [Online]. Available:
  \url{https://philharmonic.github.io/}
\BIBentrySTDinterwordspacing

\bibitem{fan_power_2007}
\BIBentryALTinterwordspacing
X.~Fan, W.-D. Weber, and L.~A. Barroso, ``Power provisioning for a
  warehouse-sized computer,'' in \emph{Proceedings of the 34th annual
  international symposium on {Computer} architecture}, ser. {ISCA} '07.\hskip
  1em plus 0.5em minus 0.4em\relax New York, NY, USA: ACM, 2007, pp. 13--23.
\BIBentrySTDinterwordspacing

\bibitem{efron_introduction_1994}
B.~Efron and R.~J. Tibshirani, \emph{\BIBforeignlanguage{en}{An {Introduction}
  to the {Bootstrap}}}.\hskip 1em plus 0.5em minus 0.4em\relax CRC Press, May
  1994.

\end{thebibliography}

%
%
%

%

\begin{IEEEbiography}[{\includegraphics[width=1in,height=1.25in,clip,keepaspectratio]{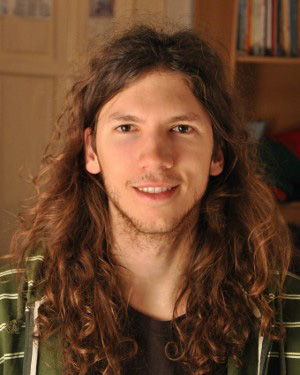}}]{Dražen Lučanin}
is a PhD student at the Vienna University of Technology,\
studying energy efficiency in cloud computing.\
Previously, he worked as an external associate at the Ruđer Bošković Institute\
on machine learning methods for forecasting financial crises.\
He graduated with a master's degree in computer science\
at the Faculty of electrical engineering and computing, University of Zagreb.\
For more information,\
please visit \url{http://www.infosys.tuwien.ac.at/staff/drazen/}
\end{IEEEbiography}

\begin{IEEEbiography}[{\includegraphics[width=1in,height=1.25in,clip,keepaspectratio]{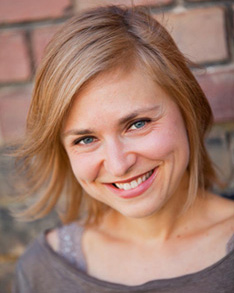}}]{Ivona Brandic}
is Assistant Professor at the Vienna University of Technology. Prior to that, she was Assistant Professor at the Department of Scientific Computing, University of Vienna. She received her PhD degree in 2007 and her venia docendi for practical computer science in 2013, both from Vienna University of Technology. In 2011 she received the Distinguished Young Scientist Award from the Vienna University of Technology for her project on the Holistic Energy Efficient Hybrid Clouds. She published more than 50 scientific journal, magazine and conference publications and she co-authored a text-book on federated and self-manageable Cloud infrastructures. For more information, please visit \url{http://www.infosys.tuwien.ac.at/staff/ivona/}
\end{IEEEbiography}



%
%

\end{document}